%% file: main.tex
\documentclass[aip, rsi,
 amsmath,amssymb,
 reprint,
]{revtex4-1}

\usepackage{graphicx}
\usepackage{dcolumn}
\usepackage{bm}

\usepackage[utf8]{inputenc}
\usepackage[T1]{fontenc}
\usepackage{mathptmx}
\usepackage{etoolbox}
\usepackage{textcomp}
\usepackage{comment}
\usepackage{booktabs}
\usepackage{subcaption}

\usepackage{tabularx}
\newcolumntype{Y}{>{\raggedright\arraybackslash}X}
\usepackage{booktabs}
\usepackage{wasysym}

\usepackage{xcolor}

\usepackage{array}
\makeatletter
\def\@email#1#2{%
 \endgroup
 \patchcmd{\titleblock@produce}
  {\frontmatter@RRAPformat}
  {\frontmatter@RRAPformat{\produce@RRAP{*#1\href{mailto:#2}{#2}}}\frontmatter@RRAPformat}
  {}{}
}%
\makeatother
\begin{document}

\preprint{AIP/123-QED}

\title[Analysis and Uncertainty Quantification of Thermal Transport Measurements through Bayesian Parameter Estimation]{Analysis and Uncertainty Quantification of Thermal Transport Measurements through Bayesian Parameter Estimation}
\author{J. Drew}
\author{S. Godse}
\author{Y. Liang}
\author{A. Pathak}
\author{J. A. Malen}
\author{R. C. Kurchin}
 \email{rkurchin@cmu.edu}
\affiliation{ 
Carnegie Mellon University, Pittsburgh, PA, USA
}

\date{\today}

\begin{abstract}
The thermal transport community is increasingly interested in rigorous uncertainty quantification (UQ) of their measurements. In this work, we argue that Bayesian parameter estimation (BPE) represents a powerful framework for both analysis/fitting and UQ. We provide a detailed walkthrough of the technique (including code to duplicate our results) and example analysis based on measuring the thermal conductance of a gold/sapphire interface with FDTR. Comparisons are made against traditional analysis/UQ techniques adopted by the thermal transport community. Notable advantages of BPE include the interpretability of its results, including the capacity to indicate incorrect input assumptions, as well as a way to balance overall goodness of fit against prior knowledge of feasible parameter values. In some cases, incorporating this additional information can affect not only the magnitude of error bars but the inferred values themselves. 
\end{abstract}

\maketitle

\section{Introduction}
\label{intro}

Accurate thermal property measurements are crucial for the design of many systems, as well as for material selection as next-generation materials are discovered and evaluated. In many applications, thermal management can be performance-limiting, particularly in computing, aerospace, and energy materials.\cite{prakash_transistor_2021, barako_integrated_2018, nobrega_review_2024, kantharaj_heat_2019, moreno_electric-drive_2021, warzoha_applications_2021} 

However, most thermal property measurement techniques require an underlying physical model to translate the direct measurement (e.g. temperature rise) into the associated thermal property value (e.g. thermal conductivity). These models will typically involve many system variables, such as those related to the energy input, the heat capacity and thermal conductivity of the associated materials, and the length scales of the sample system. As a result, when determining a thermal property, many system properties must already be known quantitatively (these will be referred to as ``externally defined inputs'' in this work). 

For each externally defined input, there is some level of uncertainty in its exact value, and its individual uncertainty will propagate through to the resulting uncertainty in the desired thermal property. In some cases, conclusions about thermal properties of interest can be quite sensitive to comparatively small uncertainties in the inputs. The resulting uncertainty in thermal property measurements can complicate precise system design, and in some cases can obscure conclusions about the best choice of material from a handful of options.  

Although various approaches exist to account for this uncertainty and quantify its impact on the resulting property measurement (several examples are outlined in more detail in Sec.~\ref{trad_uq}), there are shortcomings associated with many traditional techniques. In judging the benefits and drawbacks of various techniques, it can be helpful to list some desiderata for any approach to model fitting and uncertainty quantification (UQ):

\begin{itemize}
\item \textbf{Considering alternative input values.} For externally defined inputs, there is some uncertainty in their precise value. Some techniques merely attempt to approximate this uncertainty post hoc. However, it is more realistic to examine how the system would behave if these inputs had different values than their mean.
\item \textbf{Capturing variable correlation.} Some techniques assume all input uncertainties are independent of each other. In practice, the value of one input that best predicts some measurements will often be correlated with the value of another input; thus, uncertainties may also be correlated. This can be investigated by varying multiple inputs simultaneously.
\item \textbf{Considering resulting fit.} For techniques that examine alternative values of the externally defined inputs, the goodness of fit should affect the relative likelihood of the solution set. 
\item \textbf{Ease of implementation.} Analysis/UQ  techniques should not be operationally challenging to their specific experimental community.
\item \textbf{Computational cost.} Techniques should not be prohibitively costly in computational power, as they will often be done without access to high-performance computing resources.
\item \textbf{Interpretability of results.} Beyond the best-fit value and associated uncertainty of an estimate, there are many additional insights that can be gleaned. This can include the relative contributions of each input to the resulting uncertainty, correlations between the inputs, and the relative goodness of fit of the best-fit value vs. other possible values. While these insights are always in principle accessible, they emerge much more naturally from some techniques than others.
\item \textbf{Incorporating prior knowledge.} An experienced researcher is often aware of a range of values that are physically reasonable, and consistent with prior work in the field. This can subjectively affirm or call into doubt a given set of results; however, it is possible to explicitly incorporate this external context directly into some analyses.
\end{itemize}

In this work, we argue that Bayesian parameter estimation (BPE) is a powerful alternative to traditional uncertainty quantification approaches, such as root sum of squares aggregation of single parameter uncertainties or Monte Carlo methods, and compare them across the above list of criteria. BPE leverages Bayes' theorem to output a probability landscape of possible values of the parameters of interest, and can do so while exploring a multi-variable range of inputs. 

Herein, we step through an application of BPE used to determine the interfacial thermal conductance of a gold/sapphire interface, and examine the results compared to those obtained by traditional methods. Measurements are obtained using frequency domain thermoreflectance (FDTR), an optical technique typically used in multi-layered systems, which require knowledge of many externally defined inputs in their physical model. However, this method could easily be adapted to time domain thermoreflectance (TDTR), the 3$\omega$ method, or any other measurement with a well-defined physical model.

\section{Background}

\subsection{Frequency Domain Thermoreflectance (FDTR)}
The thermal property measurements analyzed in this work were conducted using frequency domain thermoreflectance (FDTR), a laser-based, non-contact optical technique.\cite{schmidt_frequency-domain_2009, malen_optical_2011, regner_instrumentation_2013, kirsch_instrumentation_2024} In FDTR, two lasers are co-aligned and focused onto the sample of interest: a ``pump'' laser, which heats the sample, and a ``probe'' laser, which is reflected off the surface of the material. These lasers are chosen to be at different wavelengths such that the pump laser is absorbed by the surface and the probe laser is reflected, to as great a degree as possible. A thin transducer layer with high coefficient of thermoreflectance at the probe wavelength is deposited on top of the sample (e.g. 75 nm of Au for our probe wavelength of 532 nm), to allow for measurement of temperature based on the intensity of the reflected probe laser.  The use of a transducer enables interrogation of a material's thermal properties independent of its native optical properties.

The pump laser's intensity is modulated to create an oscillating heating profile at the top surface of the material. The probe laser then measures the resulting oscillations in the temperature of the material. The pump and probe lasers can then be compared at the modulation frequency, and the phase lag between them represents the phase lag between the heat flux and the surface temperature. This process is repeated over frequencies that typically range from $10^5$--$10^7$ Hz. The phase lag as a function of frequency is then fitted with an analytical model to determine information about thermal properties of the material.

For the analytical model describing the outcome of these measurements, we follow a solution to the heat diffusion equation outlined by Cahill.\cite{cahill_analysis_2004} In that work, the temperature profile of a layered material is derived in cylindrical coordinates, which describes our case of co-aligned Gaussian lasers for heating and measurement: This model predicts phase lag, subject to input parameters of laser radius, and the thickness, thermal conductivity, and volumetric heat capacity of each layer. It can also incorporate interface conductances between the layers. As will be discussed in the analysis, uncertainty in each of these input parameters has an associated impact on our confidence in the resulting fit of our model.

\subsection{Least-Squares Regression}
Least-squares regression (LSR) is the most widely-used approach to fit a model to data. The core idea is that, given a set of experimentally measured data, and some mathematical model that describes the system, we can determine properties of the system itself. This is accomplished by finding the values of the input parameters that result in the modeled data that most closely matches the measured data. We take this closer agreement to signify more accurate representation of the system.

The name ``least-squares regression'' refers to how we define the level of agreement between the model and the experimental measurements: in the case of FDTR, for every measurement taken at a given pump laser modulation frequency $f_i$, we first calculate the difference between the measured experimental phase lag $\theta_i$ and  the predicted model value $g(f_i,\vec p)$ (where $\vec p$ represents the model parameters to be estimated). These differences are referred to as the residuals, and can be computed at any point $\vec p$ in the parameter space. Then, these residuals are squared, which symmetrizes positive vs. negative deviations, and also puts a greater penalty on measurements that are farther from the predicted value:
 \begin{equation}
\label{sse}
    S(\vec p; \{\theta,f\}) = \sum_{i=1}^n \left| \theta_i - g(f_i; \vec p)\right|^2.
\end{equation}
In LSR, the model that has the lowest sum of these squared residuals $S$ (or ``sum of squared error'', SSE) is considered to be the best-fitting model.
It is of note that in this framework, when the model with the lowest SSE is found, the associated fitting parameter’s values $\vec p$ are declared the ``inferred values''. These are the parameter values that  will be output when using an automatic fitting routine, such as MATLAB’s \verb|nlinfit|, or python’s \verb|curvefit|. This matters, because a set of inferred parameter values \textit{will} be reported, regardless of how much error there is between the model and the experiments. This occurs because that set of model parameters is the best one available, even if it is simply the best of many bad options. This approach can fail to capture the uncertainties and non-idealities associated with a fit, as will be discussed in later sections. 

\subsection{Bayesian Parameter Estimation (BPE)}
\label{sec:BPE_background}
Bayesian parameter estimation approaches model fitting via a different framework than LSR. Rather than finding only the minimum SSE value of the parameters and reporting that position as the sole inferred set of values, BPE focuses on how \textit{good} of a fit a given set of parameters is relative to others. 
It considers (and reports out) the entire landscape of the combinations of parameter values, rather than focusing on a single local minimum of SSE, allowing for direct examination of multi-variable behavior in the space of the model. It can rigorously and interpretably incorporate prior knowledge about the system, such as data from other sources about fitting parameters, without imposing rigid constraints on their values. And finally, it explicitly incorporates the uncertainty associated with the system (both of the measurements and of the externally defined inputs), in order to convert the results from simply SSE’s into actual probabilities of a given inferred parameter value being correct.

It is helpful to begin by discussing the mapping of the parameter landscape. Rather than LSR’s approach to finding only the SSE-minimizing set of parameters, BPE will evaluate every combination of possible values for the parameters, and generate model data for each. In practice, due to computational cost, discrete parameter values are sampled (in this work, they will be on a grid, though other schemes are possible), and some reasonable lower and upper bounds are placed on the possible values. Ideally, these should be chosen such that the bounds and the granularity of the combinations do not meaningfully affect the conclusions drawn (discussed in more detail below). It is also worth noting that we do not need to consider all possible values of a parameter equally likely in cases where we have some external knowledge, and this will also be discussed in detail.

Because BPE explores the full parameter space, it also naturally encapsulates any multi-variable correlation structure present in the system. This can be useful in cases where the output of a model is not uniquely constrained by one parameter. A relevant example of this exists in FDTR, where a layer’s thickness and its thermal conductivity can have many pairwise combinations with near-identical thermal resistance (a thick layer with high thermal conductivity may have the same thermal resistance as a thinner layer with low thermal conductivity). BPE has been demonstrated to identify this multivariable behavior in prior work when there was poor resolution in the individual parameters,\cite{brandt2017rapid} and indeed we have seen this thickness/thermal conductivity tradeoff appear in the system studied in this work.

A crucial and potentially unfamiliar aspect of BPE is the likelihood function $\mathcal{L}(\vec p)$, which relates the residual of a measurement to an associated probability of the set of parameters $\vec p$ which resulted in that given residual. In this case, we use a Gaussian likelihood function, which is suitable for most systems as measurement noise tends to be Gaussian~\cite{constable1988parameter}:
 \begin{eqnarray}
    \label{eq:gauss_lkl}
    \mathcal{L}(\vec p; \{\theta,f\}) &=& \exp\left(-\sum_{i=1}^n \frac{\left(\theta_i-g(f_i;\vec p)\right)^2}{2\sigma_{\text{meas,}i}^2}\right)\\
    &=&
    \label{eq:guass_lkl_homosked}\exp\left(-\frac{S(\vec p; \{\theta,f\})}{2\sigma_{\text{meas}}^2}\right).
\end{eqnarray}

Here, as above, $(f_i, \theta_i)$ and $g(f_i,\vec p)$ represent a phase lag measurement $\theta_i$ at some frequency $f_i$, and the model’s predicted result for that same measurement with parameters $\vec p$, respectively. An intuitive interpretation of a single term in this likelihood function is that it tells us ``how likely we would have been'' to measure some other phase lag $\theta$ at the same frequency $f$. For Gaussian noise with variance $\sigma_{\text{meas,}i}^2$, the uncertainty $\sigma_{\text{meas,}i}$ thus provides the natural ``decay scale'' for this quantity.
If $\sigma_{\text{meas,}i}$ is the same value $\sigma_{\text{meas}}$ for all data points (this trait is referred to as homoskedasticity in statistical parlance), then Equation~\ref{eq:guass_lkl_homosked} holds, and the likelihood function can be directly computed from the SSE map over the parameter space (see Sec.~\ref{msemap}).

In this work, FDTR phase lag data ($\theta_i$) were measured at multiple modulation frequencies ($f_i$). At each frequency, these phase lag measurements were time-averaged using a lock-in amplifier, and the standard deviation of those measurements was used as $\sigma_{\text{meas,}i}$. This is consistent with the choice to use the Gaussian likelihood to convert SSE to probability; when we take multiple measurements, we expect that the measurements should all correspond to a single underlying true value, and yet due to unavoidable noise in our system, we obtain some spread in the measurements.

The direct inclusion of this uncertainty in BPE is powerful for two reasons. The first is that it naturally scales the confidence in our prediction according to the level of variability of the measurement. The second is that this uncertainty can vary from measurement to measurement (heteroskedasticity). This allows measurements with high variability to have less impact on the curve fitting process than measurements in which there is a greater degree of certainty in the value recorded.

It should be explicitly stated that at the end of this procedure, BPE will normalize the probability of all considered options, making it sensitive to the \textit{relative} fit of the model (i.e. how much lower the SSE of one result is than the other possible results). This means it is possible that BPE can report a result as highly likely even if the absolute magnitude of the SSE is high, so long as the SSE is much lower than all other evaluated options. However, this typically will only occur in cases with an inadequate model or poor evaluated ranges of the input parameters, and due to the interpretability of BPE results (explored in more detail below), these signals of overall poor fit can often be detected and corrected. Examples of this are discussed in Sec. \ref{sec:biased_prior} and S1. 

Finally, there is the capability to incorporate externally known information. In BPE, we evaluate wide ranges of possible parameter values. However, we may know that certain parameter values are incredibly unlikely, based on pre-existing and independent information (such as the reported literature values or separate measurements which make up our externally defined inputs). BPE can account for this, making it so that the possible values of a parameter are not all equally likely, even before we begin analyzing our new measurements. This is done through the use of a ``prior'', a term in Bayesian statistics that describes a probability distribution that acts as a weighting factor corresponding to our preexisting assumptions about the system. This prior and its effect on the inference will be evaluated in greater detail in Sec.~\ref{sec:multiparam}. The final result of BPE (the \emph{posterior} distribution $\mathcal{P}$) is then given as (with normalization factor omitted):
\begin{equation}
    \mathcal{P}(\vec p; \{\theta,f\})\propto \mathcal{L}(\vec p; \{\theta,f\})\pi(\vec p),
\end{equation}
where $\mathcal{L}$ represents the likelihood as defined in Eq.~\ref{eq:gauss_lkl}, and $\pi$ represents the prior. 

\subsection{Traditional Uncertainty Quantification Approaches}
\label{trad_uq}

Several methods have been used for uncertainty quantification in the thermal property measurement community with varying levels of complexity and sophistication. We will briefly outline some notable examples here and return in Sec.~\ref{other_uq} to make more explicit comparisons to BPE.\footnote{A subtle but important point to bear in mind is that the ``error bar'' resulting from a Bayesian analysis is formally a \emph{credible} interval, while that resulting from a traditional statistical analysis such as LSR is a \emph{confidence} interval. In the case of a uniform prior (i.e. not weighting any parameter values above others), these can be thought of interchangeably; however, formally, in the case of the Gaussian priors that will also be investigated in this work, these intervals quantify different things, the former related to the final BPE result (aka the posterior distribution) and the latter more directly related to the likelihood. In this work, we will generally refer simply to ``uncertainty'' or ``error bar'' to encompass both ideas.}

\subsubsection{Pythagorean Uncertainty Sum/RSS}
\label{RSS_method}
An intuitive approach is to evaluate the impact of the uncertainty in each parameter one at a time, and then aggregate these individual impacts to determine an overall impact, or overall uncertainty in the final value. This has been done in prior work measuring thermal properties, \cite{malen_optical_2011} and is expressed in the equation below:

\begin{equation}
\begin{aligned}
\Delta G &= \sqrt{ \sum_j \left( \Delta G_j \right)^2 } 
\quad\text{and}\quad
\Delta G_j = \frac{\partial G}{\partial j} \Delta j .
\label{eq:RSS}
\end{aligned}
\end{equation}

In Eq.~\ref{eq:RSS}, $j$ indexes each externally defined input parameter, and $\Delta G_j$ represents the change in the resulting interface thermal conductance when an input parameter is shifted by $\pm 2\sigma_{\text{param}}$, representing a 95\% confidence interval. This approach will be referred to as ``root-sum-of-squares'' (RSS) uncertainty quantification, in regards to how it aggregates the uncertainties. 
This Pythagorean method of aggregating the uncertainties (squaring before combining) arises from the fact that when summing random variables, the variances add linearly. However,  this approach also assumes that the associated covariance term is zero; that is, the RSS approach ignores relationships between input variables, both in its method of exploring the system response, and its calculation of the resulting uncertainty.


\subsubsection{Approaches to Incorporate Covariance}
\label{sec:covariance}
One can adopt more sophisticated approaches to account for these covariance effects, albeit with a corresponding increase in the analytical burden. Ref.~\citenum{yang_uncertainty_2016} offers a detailed analytical approach that can incorporate uncertainties in the experimental measurements and externally defined input parameters, as well as capturing correlations. However, this approach requires access to higher-order derivatives of the system behavior in order to obtain analytical results, which in turn necessitates either detailed numerical analysis (which would usually only be performed about the lowest-SSE point) or derivation by hand of the associated Jacobian matrix, which is not always feasible. 
This approach can also experience challenges in the case of no unique minimum error in the parameter space, or very shallow minima.

\subsubsection{Monte Carlo}
Monte Carlo (MC) methods are another popular approach that, like BPE, enable exploration of the full range of system behavior without access to analytical derivatives of the model.\cite{bougherThermalBoundaryResistance2016b} Large numbers of simulations with slightly perturbed parameters (whose values are drawn from distributions representing the uncertainty in these externally defined inputs) are conducted, and the results are aggregated into a distribution, e.g. of thermal conductivity values.

MC methods typically provide equal weight to every randomized run, relying on the relative frequency of an outcome as a measure of its probability. However, in doing so, the MSE of one simulation's fit relative to another is not directly considered.

\subsubsection{MSE Mapping}
\label{msemap}
The probability distribution obtained by BPE is also closely related to the MSE map, which plots mean squared error $M$ (equal to the sum of squared errors $S$ divided by the number of measurements $n$) across the parameter space of the model. 
In particular, in the homoskedastic (uniform uncertainty) case, they can be directly transformed using the second equality of Eq.~\ref{eq:gauss_lkl}. This is a manifestation of the fact that LSR is a maximum likelihood estimator. Nevertheless, MSE mapping does not incorporate a method to introduce priors.  

BPE can also help provide an interpretation to a common ``rule of thumb'' often adopted in assessing UQ from MSE maps, namely that a 95\% confidence interval is bounded by the contour of MSE equal to twice the minimum value.\cite{WangCahillUltralowThermalConductivityFullerene} Under the assumption of a ``good'' model (that is, where residuals are dominated by measurement noise as opposed to model misspecification, i.e. the model not adequately describing the system under study), the minimum MSE value (corresponding to the maximum value of the posterior) should occur at a location $\vec p_{\text{max}}$ where $M(\vec p_{\text{max}})\equiv M_{\text{min}}\approx\sigma_{\text{meas}}^2$ (equivalently, $\sqrt{M_{\text{min}}}\approx\sigma_{\text{meas}}$). That is, the deviation of the measured data from the ``best'' model should be at the scale of the noise in the data. 
While the exact appropriate ``multiplier'' on the minimum MSE (i.e. two or something else) would depend upon the number of measurements taken and details of the geometry of the parameter space, the BPE framing makes clear that the ratio of the MSE to the variance in the measurement is the relevant dimensionless scale here. 

\section{Methods}
\subsection{Sample Details}
\label{sample}

To demonstrate the BPE approach, we measure the unknown interfacial thermal conductance ($G$) between a gold transducer and a sapphire substrate using FDTR. The sample used for this analysis was a thin film of gold deposited on a sapphire substrate. Two-inch round, double-side polished prime grade $\langle0001\rangle$-oriented sapphire with a thickness of $330$ \textmu m was purchased from University Wafer Inc. A gold layer with a target thickness of $75$ nm was deposited using electron beam evaporation, under a vacuum of $2.5\times10^{-7}$ torr. Thickness of the deposited gold layer was measured using a profilometer, and found to be $82 \pm 3$ nm. The uncertainty was taken as the RMS roughness. Thermal conductivity of the gold film was inferred from its electrical conductivity using the Wiedemann-Franz law and measured using the four-point-probe (4pp) technique. Multiple 4pp measurements were taken on the sample to obtain an uncertainty estimate. Layer thicknesses, their thermal properties, and associated uncertainties are reported in Table~\ref{tab:literature_vals}.

\begin{table*}[ht]
    \renewcommand{\arraystretch}{1}
    \centering
    \begin{tabularx}{\textwidth}{
        >{\raggedright\arraybackslash}l  
        *{6}{Y}                          
    }
        \hline
        Parameter & Units & Externally Defined Value & Standard Deviation $\sigma_{\text{param}}$ & Simulation Lower Bound& Simulation Upper Bound&Simulation Bin Size\\
        \hline
        Laser Spot Size & \textmu m& 3.4 & 0.1  & 2.88& 3.92&0.04\\
        Substrate Thickness & \textmu m& 330 &  5 (1\%) & -& -&-\\
        Substrate Heat Capacity\cite{bergman_fundamentals_2018} & J/kgK & 760 & 7.6 (1\%)  & -& -&-\\
        Substrate Thermal Conductivity\cite{collins_examining_2014} & W/mK & 38 & 0.95 (2.5\%)   & 33& 43&0.5\\
        Gold Thickness & nm & 82 & 3  & 67& 97&1\\
        Gold Heat Capacity\cite{bergman_fundamentals_2018} & J/kgK & 126 & 1.3 (1\%)  & -& -&-\\
        Gold Thermal Conductivity & W/mK & 235 & 5.9 (2.5\%)  & -& -&-\\
        \hline
 Interface Thermal Conductance& MW/m$^2$K& - & - & 32& 46&0.5\\
    \end{tabularx}
    \caption{Externally defined input parameters used in the heat transfer model. Values are obtained from a combination of experimental measurements, literature values,\cite{bergman_fundamentals_2018, collins_examining_2014} and supplier specifications, as outlined in Sec.~\ref{sample}. For parameters incorporated in simulation and analysis, simulation bounds and bin sizes are also provided. For Gaussian priors, the ``known" value serves as the mean, and the uncertainty serves as the standard deviation $\sigma_{\text{param}}$. In all cases, priors were zero outside of bounds.}
    \label{tab:literature_vals}
\end{table*}

For a single frequency scan, one specific location of the sample was measured by FDTR. The general setup of the instrumentation is described in Ref.~\citenum{regner_broadband_2013}. In each scan, 25 different modulation frequencies were chosen, logarithmically spaced from $100$ kHz to $5$ MHz, and the resulting phase lag between the pump and probe lasers was measured at each frequency. The periodic signal was detected using a lock-in amplifier (Zurich Instruments HF2LI). The time constant of the lock-in is chosen as $0.3$s, within which the signal amplitude, phase and their respective standard deviations were obtained at each modulation frequency. For measurement tools which do not natively report out standard deviations, multiple measurements could be taken and their corresponding standard deviation calculated and used as $\sigma_{\text{meas}}$.

In total, 25 frequency scans were performed on the sample located on a 5 $\times$ 5 grid of locations offset by $30$ \textmu m, over a total sample area of $120$ \textmu m $\times$ $120$ \textmu m. For much of this work, the position with the median inferred value was used as our single measured location, except where it is explicitly stated that multiple measurement locations are being considered.

\subsection{BPE}
\label{sec:bayesim}

To perform the BPE analysis, the software package \verb|bayesim|~\cite{kurchin_bayesim_2019} was used. The software takes inputs for both the experimental measurements and a set of modeled data corresponding to each of these measurements. It then computes a Gaussian likelihood for each measurement and determines the relative probability of each unique combination of parameters. The uncertainty of the Gaussian likelihood was set to be the standard deviation of each experimental measurement, which were obtained from the time-averaged phase lag measurements at each frequency.

Our analysis evaluates the model at discrete values of all input parameters, rather than a truly continuous set of possible values. The uncertainty resulting from this coarseness of the model is accounted for by the \verb|bayesim| package, and considered alongside the experimental measurement uncertainty. In addition, overly coarse evaluations will be somewhat evident visually in the resulting distributions, although this is qualitative (see Fig. S1). 
The boundaries and bin sizes of the simulations are reported in Table~\ref{tab:literature_vals}.

In this work, we investigated three sets of priors for our externally defined input parameters: a \textit{uniform} prior in which all values were considered equally likely before taking measurements; a \textit{spike} prior, in which only the precise externally defined value was considered (equivalent to fixing these parameters exactly at their ``known'' values, as traditional methods do); and a \textit{Gaussian} prior, which served as a hybrid. By imposing the Gaussian prior, we have the weighting factor be equivalent to a Gaussian distribution centered at the externally defined value of our parameters, with standard deviations equal to the uncertainty in each parameter. This is shown in Fig.~\ref{fig:different_priors}. Table~\ref{tab:literature_vals} summarizes parameter bounds (applicable to all priors) as well as means (i.e. values for spike priors) and standard deviations for Gaussian priors.

\subsection{Monte Carlo}
In this work, we analyze the same data via both MC and BPE to draw comparisons between the techniques. The approach taken was to draw values for each of the three externally defined input parameters (laser spot size, substrate thermal conductivity, and gold thickness) according to the same Gaussian distributions summarized in 
Table~\ref{tab:literature_vals}. Then, LSR was used to determine the interface thermal conductance, treating the drawn values for the three externally defined parameters as fixed.
This process was repeated for 3000 iterations, with each iteration having one unique combination of the three perturbed input parameters, resulting in one inferred interface thermal conductance obtained from that combination of perturbed inputs. 

The resulting fits for each randomized set of inputs are then aggregated. This results in four distributions -- one for the inferred interface thermal conductance, and three for the perturbed values of the externally defined input parameters that correspond to that inference.

\section{BPE Results}
We begin by discussing the analysis of a single scan of phase lag vs. frequency, taken at one location on the sample. In Sec.~\ref{multiple_meas}, we will briefly explore a combined analysis of the full set of data at all 25 locations (see Sec.~\ref{sample} for more measurement details).

Within this section, we will begin in~\ref{sec:singlevar} by estimating a single variable (i.e. interface thermal conductance $G$), with spike priors on the other three parameters. Next we move to considering all four of these parameters with both uniform and Gaussian priors in~\ref{sec:multiparam}, and explore some additional advantages of BPE in Sec~\ref{sec:biased_prior}, before comparing our results to those from other analysis techniques in Sec.~\ref{other_uq} (Sec.~\ref{mc_results} reports Monte Carlo results).

\subsection{Estimating a Single Variable}
\label{sec:singlevar}
\begin{figure*}
    \centering
    \includegraphics[width=0.99\linewidth]{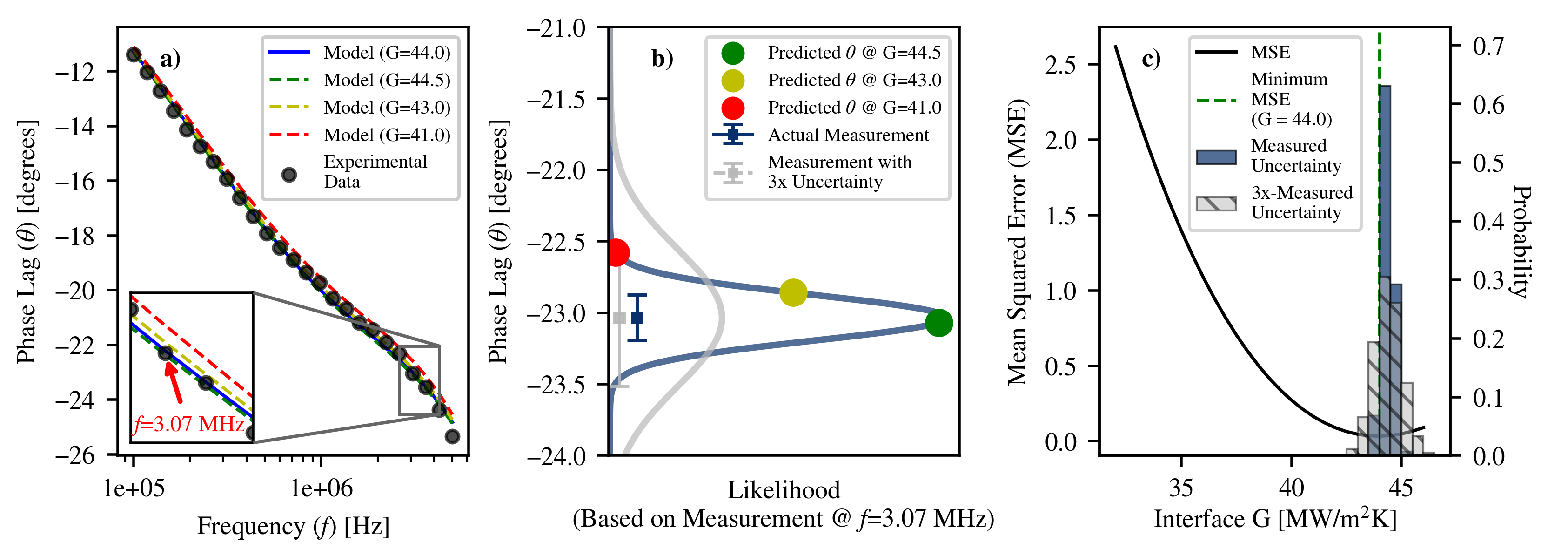}
    \caption{Example workflow for inferring solely the interface thermal conductance $G$ using BPE. All other model parameters are fixed at their externally defined values. In (a), the experimentally measured phase lag data (black) are compared to  the overall best-fit model (solid blue), as well as 3 other possible interface conductance values chosen for demonstration purposes (dashed green, yellow, and red). In (b), a single measurement (arbitrarily chosen at 3.07 MHz) is examined to demonstrate how likelihoods of a given model are determined according to their location relative to a normal distribution, with a mean and standard deviation ($\sigma_{\text{meas,}i}$ in Eq.~\ref{eq:gauss_lkl}) corresponding to the measurement. The actual measurement (dark blue) is shown alongside  a hypothetical measurement with the same mean, but an uncertainty $3\times$ greater than what was measured (gray), for demonstration purposes. Again, three possible values of the interface conductance are shown, with decreasing likelihood as they deviate more greatly from the measured value (green, yellow, red). In (c), phase lag measurements at 25 frequencies are aggregated, the resulting MSE (solid black) corresponding to different possible values of the interface conductance is shown, and the minimum MSE point is designated (dashed green, vertical). In addition, the probability distributions generated by BPE are shown, both for the actual measurements (blue) and the hypothetical set of measurements with $3\times$ the magnitude of the uncertainties (gray).}
    \label{fig:single_measurement_infer}
\end{figure*}

To infer a single parameter, a BPE framework operates almost identically to classical LSR. First, measurements are taken on the system, which in this work is a series of phase lag measurements taken from the sample at different modulation frequencies, as shown in Fig. \ref{fig:single_measurement_infer}a. This set of measurements can be used to calculate an SSE and MSE for different possible values of the inferred variable. In the LSR approach, the value of the inferred variable with lowest SSE is then reported as the inferred variable value, and the landscape of SSE vs. parameter landscape is discarded.

However, in a BPE framework, we retain the information associated with other possible values of the inferred variable. Rather than a single value, results are presented as a probability landscape. To achieve this conversion, we use a Gaussian likelihood function (Eq.~\ref{eq:gauss_lkl}). This calculation is depicted for a single measurement frequency in Fig.~\ref{fig:single_measurement_infer}b. A similar likelihood is calculated using the phase lag measured at each frequency, and all likelihoods for a given value of the interface thermal conductance (say, $G=44.0$ MW/m$^2$K) are then multiplied together to compute the final probability of that value (up to a normalization constant). This corresponds to each phase lag measurement being independent (analogous to the outcomes of multiple coin flips). The resulting product of the likelihood calculations are then normalized to generate the relative probability distribution in Fig.~\ref{fig:single_measurement_infer}c, along with the MSE values. 


The resulting BPE inference provides both an estimate for $G$, represented as the maximum-probability point, as well as a natural uncertainty for that estimate, corresponding to the width of the distribution. The uncertainty obtained from this probability distribution shows what other possible values of $G$ would give similarly good fits of the data, to within the experimental precision of our measurements. This is the reasoning for using the measurement uncertainty as $\sigma_{\text{meas}}$ in the Gaussian likelihood function, as mentioned in Sec. \ref{sec:BPE_background}. However, the results of using a larger value of $\sigma_{\text{meas}}$ are also shown in Fig. \ref{fig:single_measurement_infer} for demonstration purposes, and it can be seen that the resulting uncertainty in the inferred value (i.e. the width of the probability distribution) increases, but the inferred value itself (i.e. the maximum-probability point) does not change.

This single-variable estimation is a useful introduction to BPE. However, the results we obtain are not the final uncertainty with which we can report our interface thermal conductance. This is because we must also take into account uncertainties in the other inputs to the model, what we have been referring to as our ``externally defined inputs,'' which have been ignored in this one-dimensional analysis. These can correspond to either an imprecise knowledge of a singular true value, such as values of thermal properties reported in other literature, or actual underlying distributions of values, such as the variation of a layer thickness due to the roughness of the layer. 

\begin{figure}[ht]
    \centering

        \includegraphics[width=0.95\linewidth]{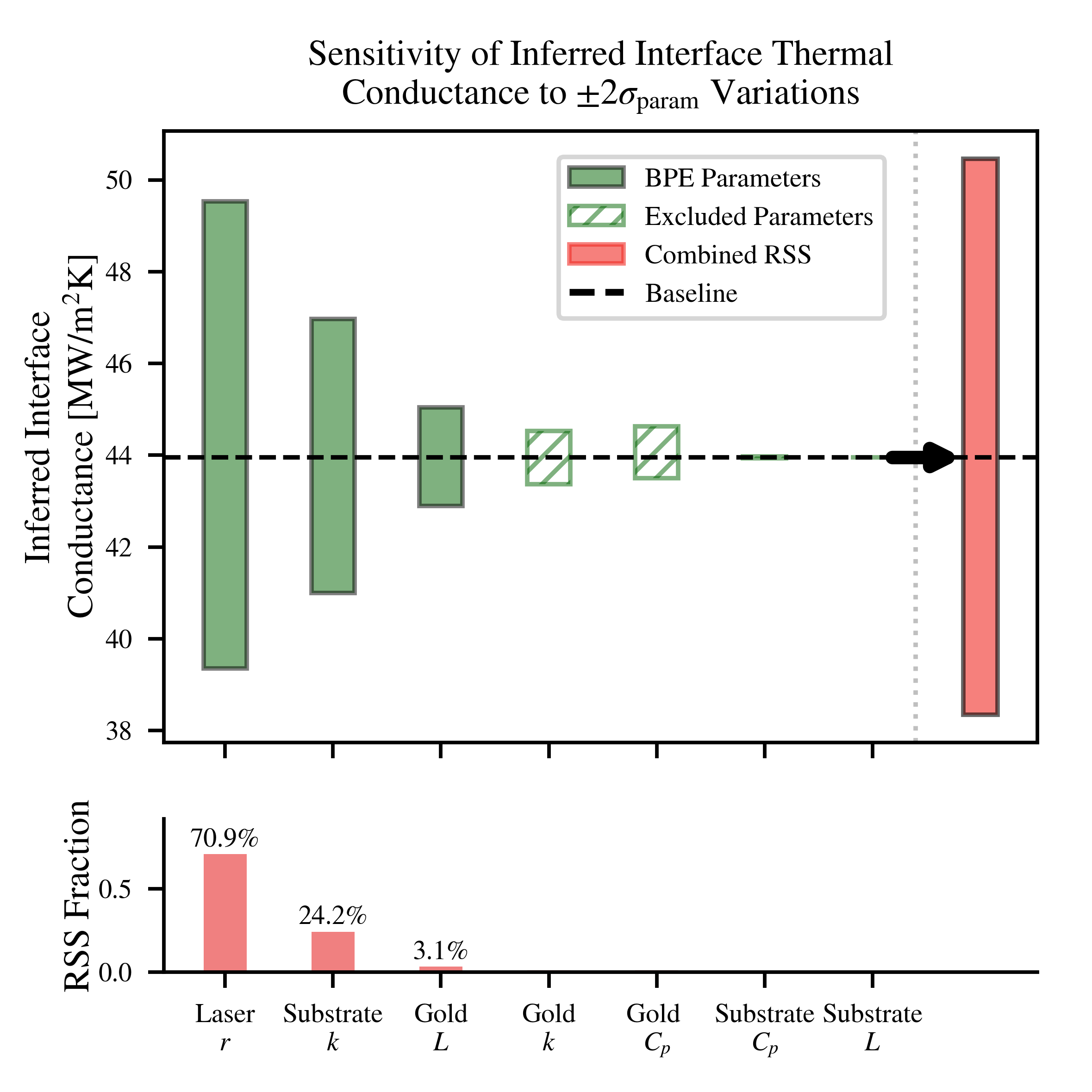}
        
    \caption{RSS uncertainty contribution analysis. The influence of each parameter's uncertainty on the measured interface thermal conductance is plotted individually. Subsequently, these individual contributions are aggregated as outlined in Eq.~\ref{eq:RSS}, and displayed on the right, in red. Each parameter's relative contribution to the aggregated uncertainty is shown below. $r$ represents the laser radius, while $k, L,$ and $C_p$ represent the thermal conductivity, thickness, and heat capacity of a layer respectively.}
\label{fig:1var_sensitivity_bars}    
\end{figure}

A common approach in the thermal transport community is to  assess each parameter individually to determine their impact on the resulting estimate, through the RSS method outlined in Sec.~\ref{RSS_method}.
The results of this analysis are shown in Fig.~\ref{fig:1var_sensitivity_bars}, where the $\pm 2\sigma_{\text{param}}$ values of each parameter from Table~\ref{tab:literature_vals} are used as inputs to the model individually, while all other parameters are held constant (for example, the inference is conducted assuming a laser spot size of 3.6 \textmu m -- i.e., 3.4\textmu m + 2 $\times$ 0.1\textmu m -- while all other inputs are held at their externally defined value). In a single-variable inference framework, the RSS total of these model uncertainties and the measurement uncertainty provide a total uncertainty in the inferred interface thermal conductance.

\subsection{Estimating Multiple Parameters}
\label{sec:multiparam}
Rather than simply adding the uncertainty associated with the externally defined input parameters after conducting an inference, it is possible to incorporate them into the analysis directly, by treating them as additional inference parameters. In a BPE framework, this entails checking every unique combination of values for each of the inference parameters. In this demonstration, 3 additional parameters are added to the inference: laser spot size, substrate thermal conductivity, and gold thickness. These are chosen both due to their large associated sensitivities within the model (as can be seen from Fig.~\ref{fig:1var_sensitivity_bars}), as well as our relative lack of confidence in their true values (see Table~\ref{tab:literature_vals}). 

\begin{figure}
    \centering
    \includegraphics[width=1\linewidth]{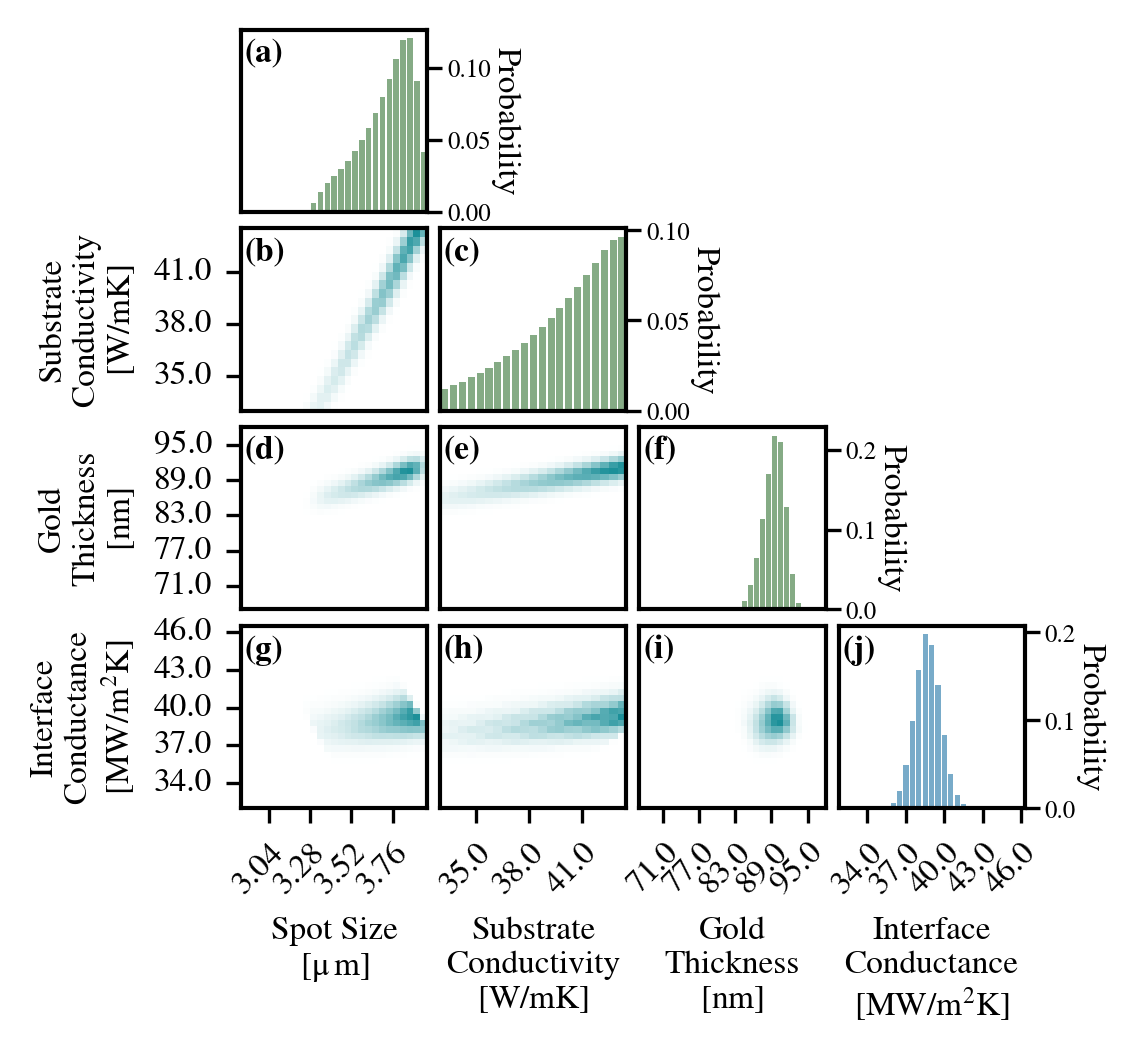}
    \caption{Resulting probabilities from the multi-parameter inference with a uniform prior. The four on-diagonal plots (a, c, f, j) represent the marginalized probability distributions of each individual parameter, with the y-axis representing the relative probability that any individual parameter value is correct. The six off-diagonal plots (b, d-e, g-i) represent probability heatmaps of each pairwise combination of parameters, showing the most likely pairwise combinations of the parameter values.}
    \label{fig:pairplot}
\end{figure}

The result of this multiparameter inference is shown in Fig.~\ref{fig:pairplot}. There are a variety of benefits to adding these additional inference parameters. One is the ability to automatically draw a good inference even in the case where the externally defined value is not precisely correct, as will be discussed in Sec.~\ref{sec:biased_prior}. Another advantage is the ability to capture multi-variable effects. One example of a pair of variables exhibiting such an effect is laser spot size and substrate thermal conductivity. We know that there is a physical relationship between how they affect the system: the laser spot size will affect how thermal energy enters the surface, while the substrate thermal conductivity will affect how thermal energy is carried away from the surface. 
In the associated two-variable marginalization (Fig.~\ref{fig:pairplot}b), the line of high-probability values of these parameters shows that a combination of large laser spot size and high substrate thermal conductivity will behave similarly to small laser spot size and low substrate thermal conductivity. This implies that while there is uncertainty in each of these parameters individually, their possible joint values are more constrained. This suggests that the inference uncertainty from these two parameters is smaller than suggested by Fig.~\ref{fig:1var_sensitivity_bars} -- that is, given better information on either one of them, the contribution of the other's uncertainty would also substantially decrease, because they are not independent. While in this particular case, this relationship could be gleaned from direct mathematical analysis of the underlying model, we emphasize that the BPE framework facilitates identification of such relationships, including in cases where direct analysis would be challenging or infeasible.

\begin{figure*}[htbp]
    \centering

    \begin{subfigure}[t]{\textwidth}
        \centering
        \includegraphics[width=\textwidth]{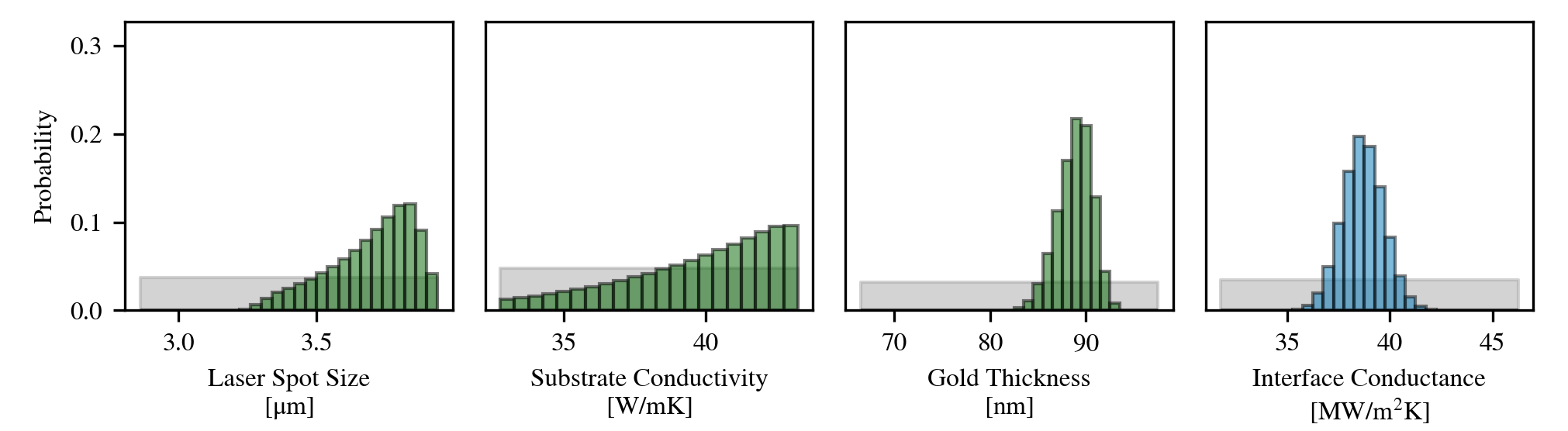}
        \caption{Results using uniform prior.}
        \label{different_priors_uniform_rev11}
    \end{subfigure}

    \vspace{1em}

    \begin{subfigure}[t]{\textwidth}
        \centering
        \includegraphics[width=\textwidth]{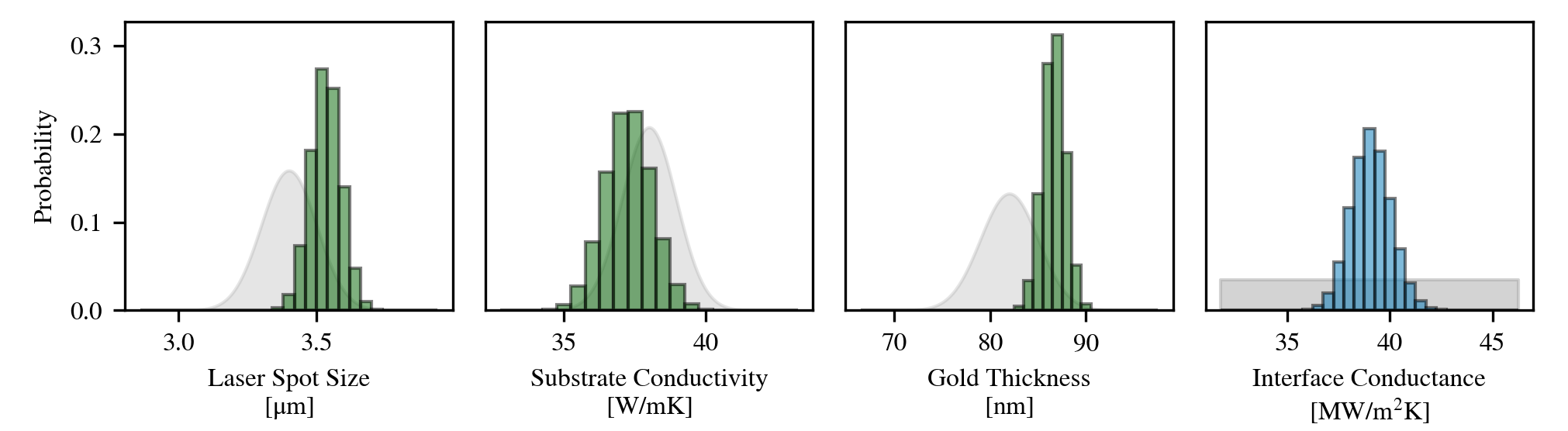}
        \caption{Results using Gaussian prior.}
        \label{fig:placeholder}
    \end{subfigure}

    \caption{Comparison of model fits obtained using uniform (a) vs. Gaussian (b) priors. Gray curves denote the associated prior used for the externally defined input parameters, while green and blue bars signify the probabilities of the resulting inferences. }
    \label{fig:different_priors}
\end{figure*}

When these externally defined input parameters are treated as inference parameters, a dilemma of sorts arises: the inference is trying to find values of these inputs that best fit the data, but there is a ``right answer,'' which should be their measured or reported values. This discrepancy arises due to the imprecision in our knowledge of the true value of these parameters (barring cases such as the roughness of a film, where there is true variation). Taken together, there are different approaches to how much emphasis we put on this outside knowledge. These all center around what we use as our \textit{prior}, or the baseline assumption we have before performing our inference. This is discussed in detail in Sec.~\ref{sec:bayesim}, through the use of \textit{spike}, \textit{uniform}, and \textit{Gaussian} priors. The impact of our choice of priors can be seen in Fig.~\ref{fig:different_priors} (the full corner plot, analogous to Fig.~\ref{fig:pairplot}, is shown in Fig.~\ref{fig:corner_plot_gaussian_prior}).

Interestingly, in this particular analysis, the uniform and Gaussian priors lead to very similar conclusions about the interface conductance. However, we can see in the distributions for the externally defined inputs that the results align much more closely with our original beliefs in the case of the Gaussian prior, whereas the results from uniform priors suggested physically implausible conclusions about these parameters.

This is relevant when we examine the MSE associated with the fits in each case, as shown in Table~\ref{tab:MSE_Penalty}. The spike priors involve absolute confidence in the externally defined values, regardless of their quality of fit. This is shown to increase the MSE of the fit by over a factor of 6 in this example. The uniform prior is a near-unconstrained error-minimization fit, although it can sometimes draw unrealistic or unphysical conclusions. The Gaussian prior is shown to be a favorable middle ground, significantly reducing the MSE as compared to absolute trust in externally defined values, while still constraining the results to reasonable values. For a comparison of the highest-probability modeled data to the experimental measurements in each case, see Fig. S4.

\begin{table}
    \centering
    \begin{tabular}{>{\centering\arraybackslash}p{0.25\linewidth}>{\centering\arraybackslash}p{0.25\linewidth}>{\centering\arraybackslash}p{0.25\linewidth}}\toprule
         Prior&  MSE& Ratio of Minimum MSE\\\midrule
         Uniform&  0.0050& 1.00\\
         Gaussian&  0.0064& 1.28\\
         Spike&  0.0334& 6.68\\ \bottomrule
    \end{tabular}
    \caption{Associated MSE values for fits conducted with each of the three choices of prior (uniform, Gaussian, and spike). As uniform priors are effectively an unconstrained MSE-minimization approach, the uniform prior MSE was considered the ``minimum MSE'', and the relative increase in MSE of the other approaches is also shown.}
    \label{tab:MSE_Penalty}
\end{table}

\subsection{Detecting Biased Prior Estimates}
\label{sec:biased_prior}
Besides correlations between variables, one of the other key benefits to BPE is the interpretability of the results. One example of this is the signal that the bins chosen are too coarse (See Fig. S1). 
However, another example from this work was in our original choice of the externally defined value for the gold thickness. 

Originally, the deposition machine setpoint of 75 nm was assumed to be the layer thickness, without independent verification after deposition. During our initial analysis, there was a clear signal that the highest-probability values of the gold thickness were at the high end of the range, even after applying a Gaussian prior set at 75 nm (See Fig. S2). 
By the classical definition of a three-sigma confidence test, this would be similar to concluding that our original hypothesis (that the gold thickness was 75 nm) was unreasonably unlikely. In addition, the experimental measurements were compared to two possible sets of modeled data: one with the gold thickness set at its original assumed value of 75 nm, and another modeled with a thicker gold layer, which was shown to have a higher probability in the Bayesian analysis. 

Following these preliminary results, the gold layer was measured with a profilometer and determined to indeed be thicker than the setpoint, with a measured value of $82\pm3$ nm. This result was much closer to the unconstrained signal from the Bayesian approach. While intuition and manual adjustment of the input parameters from their externally defined values could lead to similar conclusions by an expert, without a full exploration of the MSE landscape it is not always apparent when a better fit is possible. BPE makes the process more fail-safe and objective.


\subsection{Accounting for Sample Variability}
\label{multiple_meas}

One additional aspect of BPE as compared to traditional inference techniques is the fact that it inherently operates with distributions, rather than single reported values. Previously in this work, we have compared the ability of BPE and traditional methods to generate an overall uncertainty by combining uncertainties in measured data with uncertainties in externally defined inputs. However, that discussion focused on a single measurement (i.e. one frequency scan), from one location on a sample. In many systems, there is some nonuniformity across the sample itself, adding an additional contribution to the resulting uncertainty. 

\begin{figure}
    \centering
    \includegraphics[width=\linewidth]{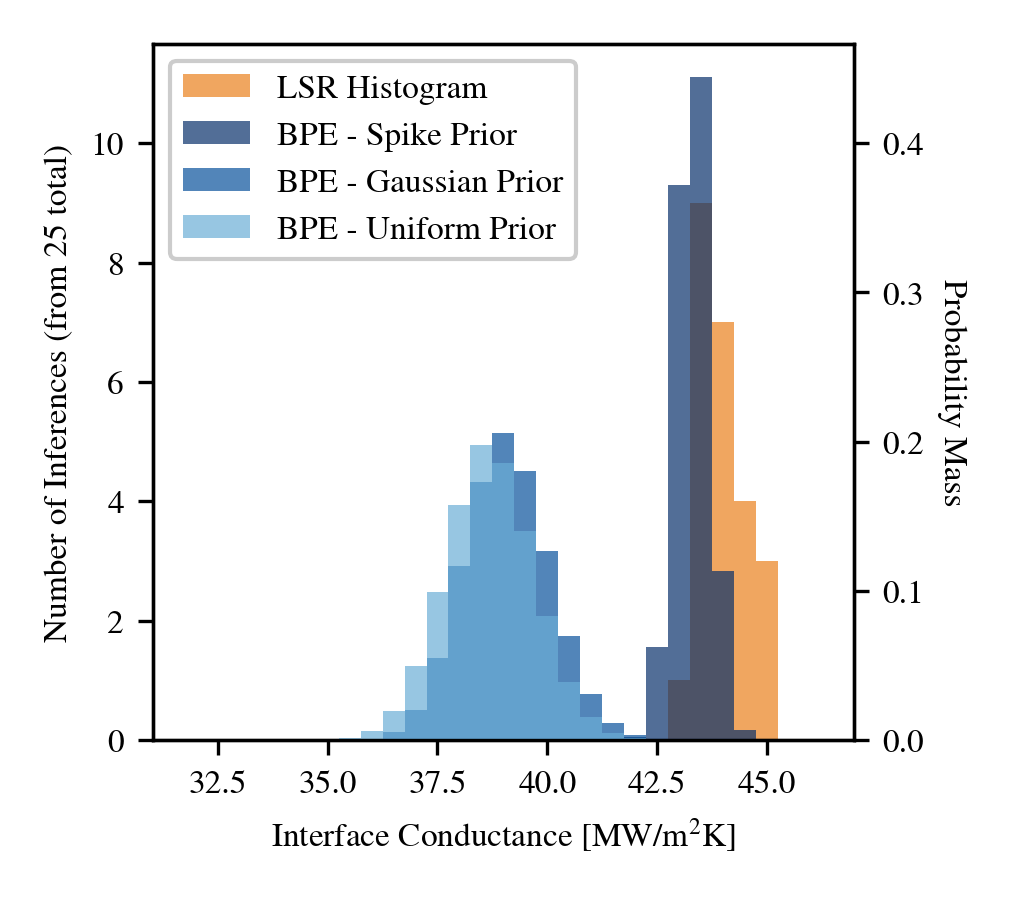}
    \caption{Resulting histogram of traditional LSR fits obtained from 25 distinct sample positions. Presented for comparison are the BPE probability distributions, associated with all three choices of prior.}
    \label{fig:multiple_meas}
\end{figure}

Taking multiple measurements and performing individual inferences in a classical LSR framework, one could create a histogram of every inferred value, and use the spread of this histogram as an uncertainty. 
In Fig.~\ref{fig:multiple_meas}, we show this histogram (in orange) for 25 measurements taken on the same sample, each in a distinct location in a 5$\times$5 grid. As shown, there is some spread in the resulting estimates, even in a classical least-squares approach. In the same figure, we show the probability distributions resulting from BPE with the same three priors as above (uniform, Gaussian, and spike), aggregated across analyses of all 25 measurement locations. 

In creating these aggregated probability distributions, a BPE analysis is performed on each location's phase lag data, and the resulting 25 probability distributions (one corresponding to each location on the sample) are then summed and renormalized. If instead the 25 positions were considered to be independent measurements of a perfectly uniform system, it would be mathematically correct to multiply these distributions, and a tighter resulting probability distribution would be expected. However, we expect some sample nonuniformity, and so the measurements taken in different locations cannot be treated as coming from a single, homogeneous system. As a result, they can be thought of as separate experiments, and there is no guarantee that the thermal behavior in one location will map perfectly onto the behavior in another.

As expected, the distribution from a spike prior is very similar to the histogram of 25 measurements. In both cases, all externally defined inputs are held at their exact defined value. There is a slight but observable shift in the highest-probability value. The only systematic difference between these approaches is that the experimental histogram is composed solely of the most-likely value from each of the 25 inferences. In place of this, the spike prior BPE result contains the entire probability distribution for each individual inference, as exemplified in Fig.~\ref{fig:single_measurement_infer}C.

The remaining two BPE distributions (uniform and Gaussian priors) both correspond to some degree of freedom for the input parameters to take values different than their externally defined values. This is shown to shift the inferred value obtained from these techniques, suggesting that this capability can alter the conclusions we draw from our inferences.

The other noticeable difference is the change in the width of the distributions. In this visualization, the experimental histogram and spike prior are both narrower. However, this is because they do not yet account for any uncertainty in the three externally defined input parameters which are incorporated into the other Bayesian analyses. When considering the overall uncertainty associated with these approaches, it can be seen that the uniform and Gaussian prior BPE distributions actually have a lower total uncertainty, as represented by the dashed line in Fig.~\ref{fig:all_bar_chart} (discussed in more detail below).

\section{Monte Carlo Results}
\label{mc_results}
\begin{figure}
    \centering
    \includegraphics[width=0.95\linewidth]{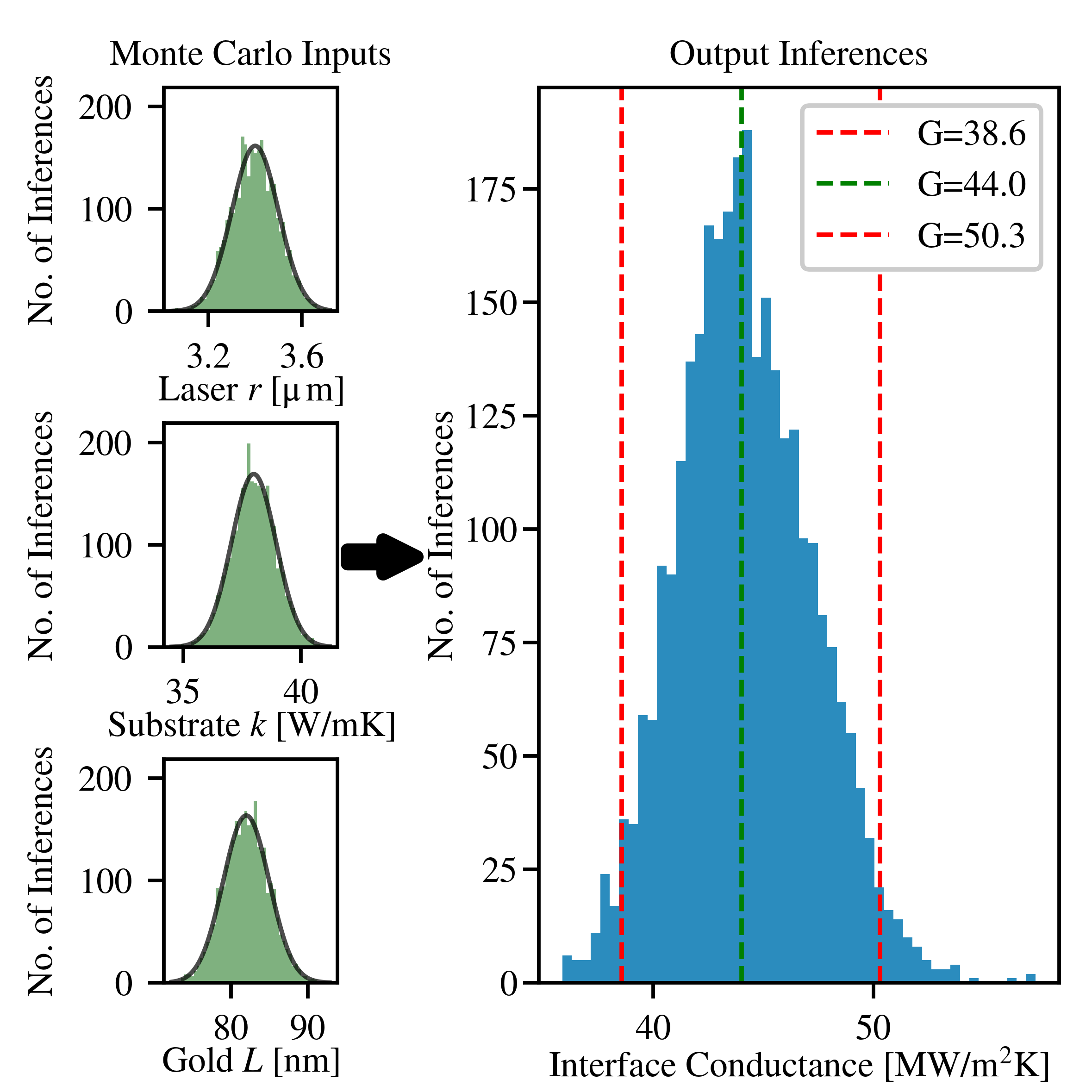}
    \caption{Monte Carlo distributions of each externally defined input parameter are shown in green, along with black curves representing the theoretical distribution from which they were sampled. The histogram of resulting interface thermal conductance values obtained from each fit are presented in blue, with the dashed lines representing the 2.5th-, 50th-, and 97.5th-percentile value of the distribution. $r$ represents the laser radius, while $k $ and $ L$ represent the thermal conductivity and thickness of a layer respectively.}
    \label{fig:mc}
\end{figure}

Results of a Monte Carlo analysis of the same data and drawing values of the externally defined inputs from the same distributions described in Table~\ref{tab:literature_vals} are shown in Fig.~\ref{fig:mc}. 
Despite the fact that MC methods do incorporate the variability of externally defined inputs directly into the analysis like BPE (and unlike RSS), the resulting distribution of interface thermal conductances is much broader than that obtained purely from BPE. This is due to the fact that in this MC approach, no consideration is given to how poorly a resulting parameter estimate actually fits the data. In BPE, the Gaussian likelihood effectively weights each inferred value proportional to how well it fits the data. Since these probabilities are all normalized, it emphasizes values which better fit the data, while still exploring these ranges of possible values.

To further explore this effect, we applied an MSE cutoff to the Monte Carlo analysis, wherein all the original steps are taken when randomly generating the externally defined input parameters, but once a model fit is conducted, it is only kept in the aggregate if it matches the experimental measurements sufficiently well (which is determined by having a sufficiently low MSE). By varying the magnitude of this MSE cutoff (See Fig. S6), 
we can see that requiring a certain goodness of fit eventually leads to a reduction in the spread of the distribution, suggesting that there is a smaller subset of externally defined input values that accurately matches the experimental behavior. Interestingly, applying this cutoff also results in an overall downward shift of the distribution of conductance values towards the BPE result.


\section{Discussion}
\label{other_uq}

\begin{figure}
    \centering
    \includegraphics[width=0.95\linewidth]{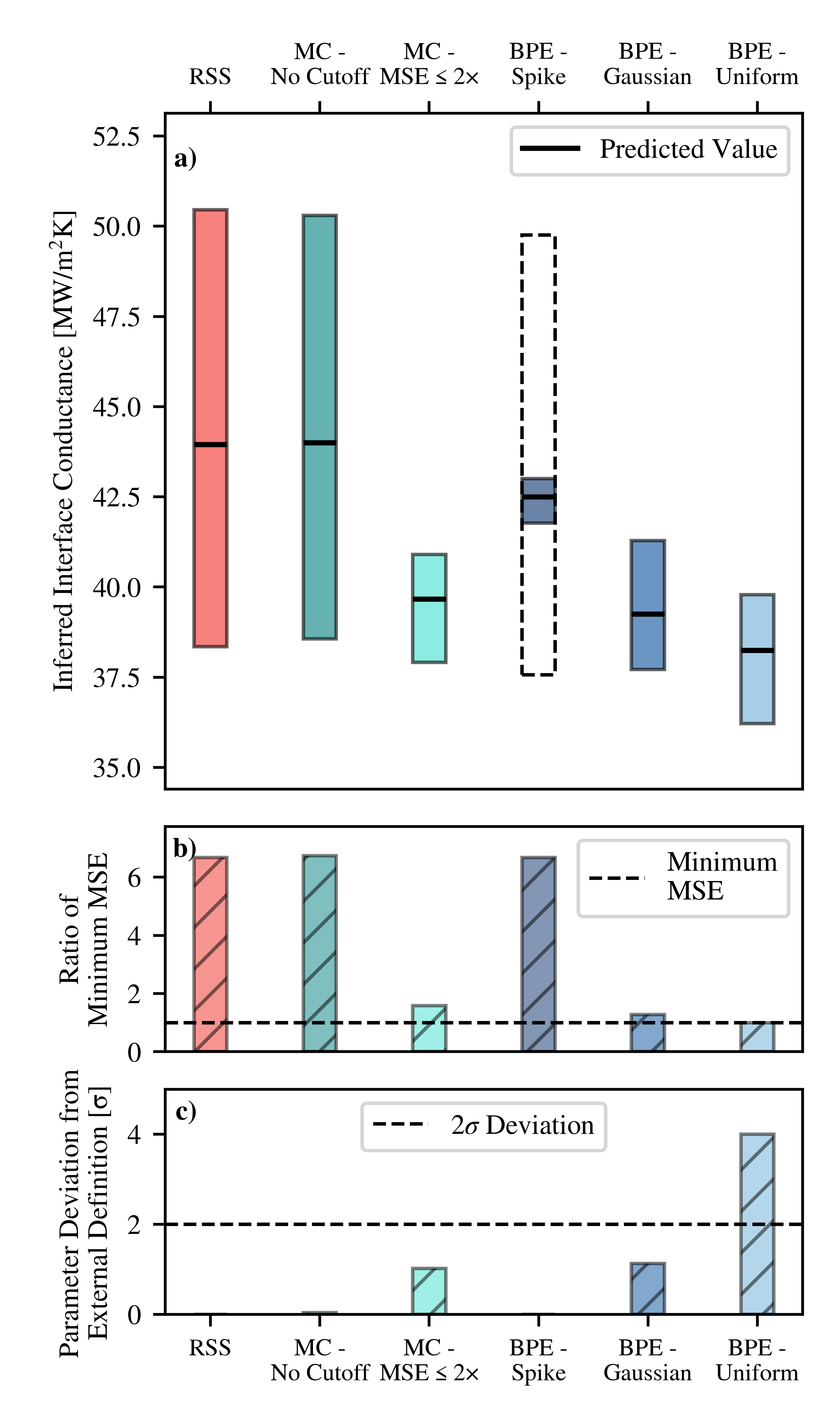}
    \caption{In (a), inferred values and uncertainty estimates of the gold/sapphire interface thermal conductance are presented, as obtained from a variety of methods. Leftmost is the range associated with a single-variable fitting's model uncertainty via RSS. In the center, ranges of the Monte Carlo distributions are shown, both without and with an MSE cutoff. Finally, on the right are the ranges of 3 BPE inferences, each with a distinct choice of prior. The dashed border surrounding the spike prior reflects that the uncertainties in externally defined inputs must be combined via RSS with the 1-variable inference, whereas in the other BPE approaches, these are incorporated into the analysis directly. In (b), the MSE of each of these techniques is shown as a ratio relative to the lowest MSE achieved by any of the techniques. In (c), the values of the three externally defined inputs (laser spot size, substrate thermal conductivity, and gold thickness) which were associated with the resulting inference were compared to their externally defined value. Plotted is the average number of standard deviations away these parameters were from their externally defined value. }
    
    \label{fig:all_bar_chart}
\end{figure}

The resulting uncertainty estimates from various techniques are compared in Fig.~\ref{fig:all_bar_chart} to demonstrate both the inferred (i.e. maximum-probability) value and associated uncertainty from each approach. 

Two key results are visible when aggregating these techniques. The first is the comparatively smaller uncertainties associated with the BPE approaches, as compared to the RSS of individual uncertainties, or the simplest MC approach. In the case of the RSS, it is possible that multi-variable correlations play a part in reducing the overall uncertainty. However, one key difference (as compared to both the RSS and the MC methods) is that BPE will incorporate a dynamic MSE weighting to the inferences (i.e. combinations providing lower MSE are weighted more heavily). 

It is particularly compelling that MSE weighting is the primary contribution to the tighter uncertainty when examining the modified MC approach, in which an MSE cutoff of $2\times$ the minimum MSE is chosen, which should achieve a somewhat similar effect as the Gaussian prior BPE approach (although with a binary include/exclude criterion, rather than dynamic, weighted incorporation of all possible values).

The other notable result is the shift of the maximum probability value of the inference in the BPE approaches (excluding the spike prior) and the MSE-weighted MC approach. These approaches explore slightly shifted values of the externally defined input parameters, and they take into account the overall MSE of the fit. Taken together, these indicate that there is a better global fit to the experimental data when these input parameters are slightly shifted from their exact externally defined values.

Returning to the list of desiderata laid out in Sec.~\ref{intro}, we can also evaluate each UQ approach described herein in light of these criteria. This analysis is summarized in Table~\ref{tab:pros_and_cons}.
The scoring of the techniques is unavoidably subjective to a degree, and where ambiguities exist, we have chosen to evaluate them in the context of the inference problem presented in this work, for equivalently precise results across the various methods.

\begin{table*}[t]
\pretolerance=10000 

\newcommand\redcirc{\ensuremath{{\color{red}\scalebox{1}{$\blacksquare$}}}}
\newcommand\yellowcirc{\ensuremath{{\color[HTML]{FFD300}\scalebox{1.1}{$\blacktriangle$}}}}
\newcommand\greencirc{\ensuremath{{\color[HTML]{007E5A }\scalebox{1.8}{$\bullet$}}}}

\centering
\begin{tabular}{|l|c|c|c|c|c|}
    \hline
    & \textbf{RSS} 
    & \textbf{Monte Carlo} 
    & \textbf{Cov. Mat.} 
    & \textbf{MSE Map} 
    & \textbf{BPE} \\
    \hline
    Considers Alternative Input Values       & \yellowcirc & \greencirc      & \yellowcirc   & \greencirc & \greencirc \\
    \hline
    Captures Variable Correlation            & \redcirc & \yellowcirc & \greencirc & \greencirc & \greencirc \\
    \hline
    Considers Resulting Fit (MSE)            & \redcirc & \redcirc    & \redcirc   & \greencirc & \greencirc \\
    \hline
    Easy to Implement                        & \greencirc & \greencirc & \yellowcirc & \greencirc & \greencirc \\
    \hline
    Computationally Cheap                    & \greencirc & \yellowcirc & \yellowcirc & \yellowcirc & \yellowcirc \\
    \hline
    Interpretable Results                    & \yellowcirc & \yellowcirc & \yellowcirc & \greencirc & \greencirc \\
    \hline
    Incorporates Prior Knowledge& \greencirc & \greencirc & \greencirc & \redcirc & \greencirc \\
    \hline
\end{tabular}

\caption{Comparison of uncertainty quantification methods common in the thermal transport community, across a list of desirable criteria. Scores are given on a qualitative basis, with \greencirc~representing strong performance in the given metric, \yellowcirc~representing moderate performance, and \redcirc~indicating poor performance or that criterion not being met at all. For some metrics, such as computational cost and ease of implementation, the performance is dependent on the physical system chosen, and the choices of the practitioner (number of MC runs, granularity of the BPE bin widths, etc.). Performance is evaluated for the system in this work, for roughly equivalent levels of precision across techniques.}
\label{tab:pros_and_cons}
\end{table*}

The RSS method is incredibly convenient to implement, and does provide some level of interpretability about the relative contributions of the individual uncertainties. However, it does not robustly explore the system behavior across the different possible input values, focusing merely at their extremes, and only for a single parameter at a time.

The MC method offers some robust system exploration, but does not provide immediately interpretable correlation information. It can provide results \textit{without} a fit-weighting for comparatively lower computational cost than a full grid exploration from BPE or an MSE map. However, the fit-weighting is shown to be highly impactful on the conclusions of an MC technique in Fig. \ref {fig:all_bar_chart}. Because MC relies on the number of occurrences as its output metric, adding a fit-weighting criteria pushes the computational cost above that of full grid-wise explorations (like BPE/MSE maps), as the extremal-value inputs must be simulated multiple times. Also, in the case where an externally defined value must be updated, MC would require an entirely new set of simulations, while BPE can simply update the prior, requiring no additional simulations.

The covariance matrix provides a great deal of information about the correlation behavior of the model inputs and the degree of their impact on the system. Its ease of implementation is particularly system-dependent, as has been discussed in Sec.~\ref{sec:covariance}. An analytical covariance matrix can be challenging to derive in some systems. A numerical investigation of discrete perturbations up or down in the values of the inputs can be easier to obtain. However, this is generally only done around the single point that matches the externally defined values of the inputs and the best-fit value of the inferred parameter, and so can suffer from the same vulnerability to incorrect externally defined values as the RSS method.

Both the MSE map and BPE effectively provide this numerical approximation of the covariance matrix, only across the entire parameter landscape as opposed to merely at the minimum MSE location (albeit likely at a higher discretization width than one would choose for a true numerical derivative). The computational cost of these two techniques is quite similar, as the amount of system modeling would be identical, with BPE only requiring a small amount of post-processing to convert SSE's to probabilities. Their performance across these metrics is only different in their ability to incorporate external information, with BPE more easily able to reveal the intersection of low-MSE regions with physically plausible values of the externally defined inputs. For reference, an MSE map of this system is included in Fig. S5.

\section{Conclusion}

BPE is a powerful tool for parameter estimation and uncertainty quantification, offering some compelling advantages over traditional approaches. These include interpretable and actionable results, e.g. in this work it was able to provide a signal that one of the initial values of an externally defined input parameter (gold layer thickness) was inaccurate. It also offers a combination of capabilities that are beneficial to model fitting, and are not typically incorporated in more traditional techniques. 

Incorporating multivariable behavior allows for a more robust approach than assuming independence in the input parameters, such as in RSS approaches. Incorporating information about the quality of fit reduces the level of uncertainty to an even greater degree, and can even result in a shift in the best-fit value, suggesting that these techniques can affect the conclusions drawn, and not just the associated error bars. Finally, BPE also allows for the incorporation of external information with adjustable levels of confidence, allowing for a significant reduction in the MSE of the resulting fit (over 5$\times$), while constraining the results to a more physically reasonable set of values than a totally unconstrained optimization.


\begin{acknowledgments}
Parts of this research were conducted using the Tartan Research Advanced Computing Environment (TRACE). The authors would like to gratefully acknowledge the College of Engineering at Carnegie Mellon University for making this shared high-performance computing resource available to its community.

This work was supported by the Army Research Office Ultra-Wide Bandgap RF Electronics Center (Award No. W911NF2220191) and the Army Research Office Multidisciplinary University Research Initiative (Award No. W911NF2310260). 
\end{acknowledgments}

\section*{Data Availability Statement}

The data and code that used in this study are openly available on GitHub at \url{https://github.com/ACME-group-CMU/FDTR-BPE-paper/tree/main}.

\section*{References}
\input{main.bbl}

\clearpage
\pagestyle{empty}
\section*{Supplementary Material}
\renewcommand{\thefigure}{S\arabic{figure}}
\setcounter{figure}{0}
\renewcommand{\thesubsection}{S\arabic{subsection}}

\subsection{Interpretability and Detecting Bias in BPE}
\label{sec:detecting_bias_SI}
\begin{figure}
    \centering
    \includegraphics[width=\linewidth]{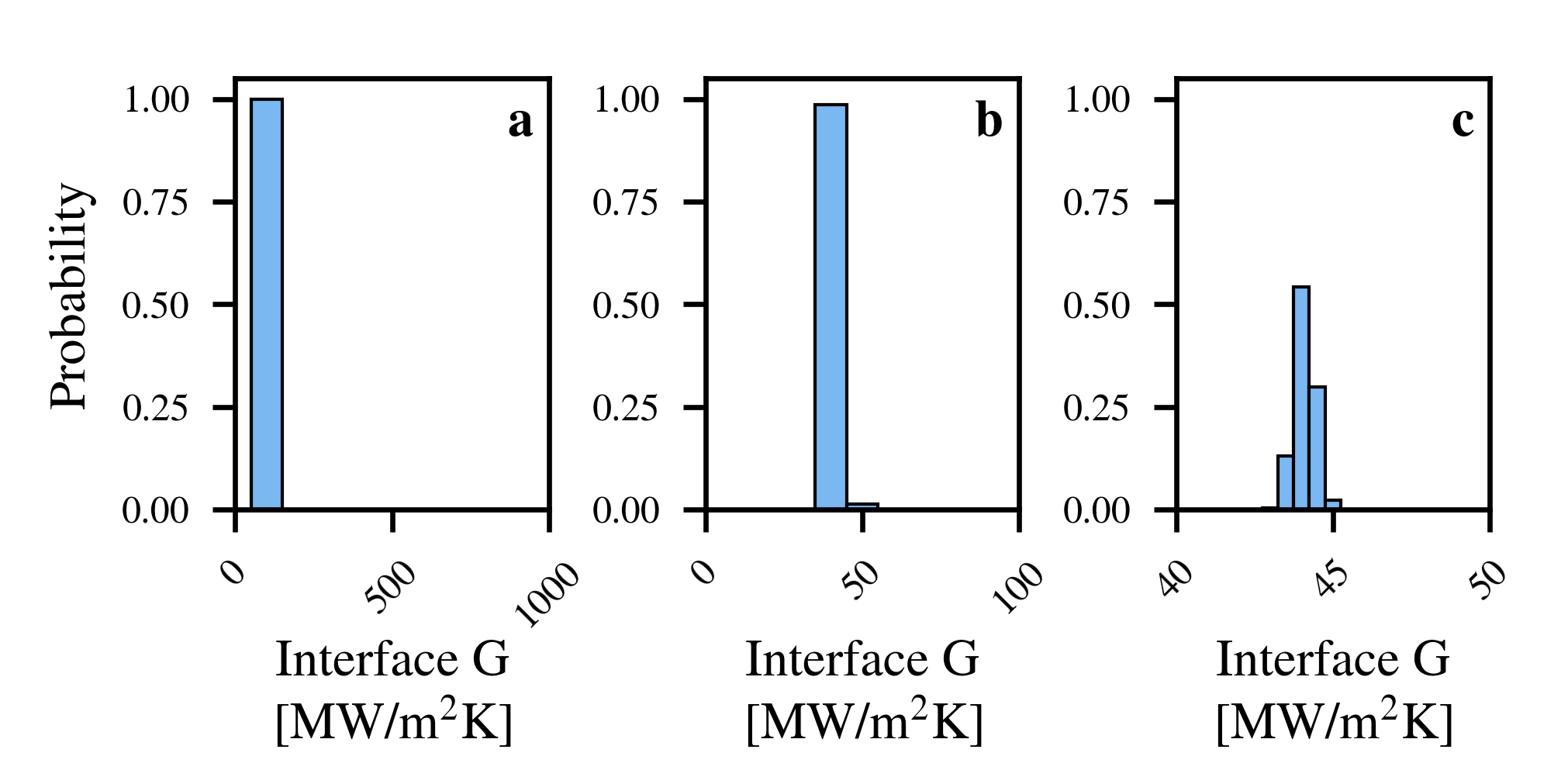}
    \caption{Posterior distributions of BPE inferences performed on identical data, with varying bin sizes and ranges. }
    \label{fig:honing_in_on_scale}
\end{figure}

As discussed in Sec.~\ref{sec:biased_prior}, a notable attribute of BPE is the interpretability of the outputs. One simple example is in the case when the appropriate scale of an input variable is unknown. Fig.~\ref{fig:honing_in_on_scale} illustrates this by way of a contrived example, where we imagine the initial range of values for $G$ was chosen far too large, such that all of the probability ``piles up'' in one bin at the edge of the range. This range is then iteratively refined until results are more informative. This highlights that even in the case of improperly chosen ranges (\ref{fig:honing_in_on_scale}a) or insufficient resolution (\ref{fig:honing_in_on_scale}b), BPE can often provide an interpretable result. This can be used to further refine the target range of an inference (\ref{fig:honing_in_on_scale}c).

\begin{figure*}[tb]
    \centering
    \includegraphics[width=\linewidth]{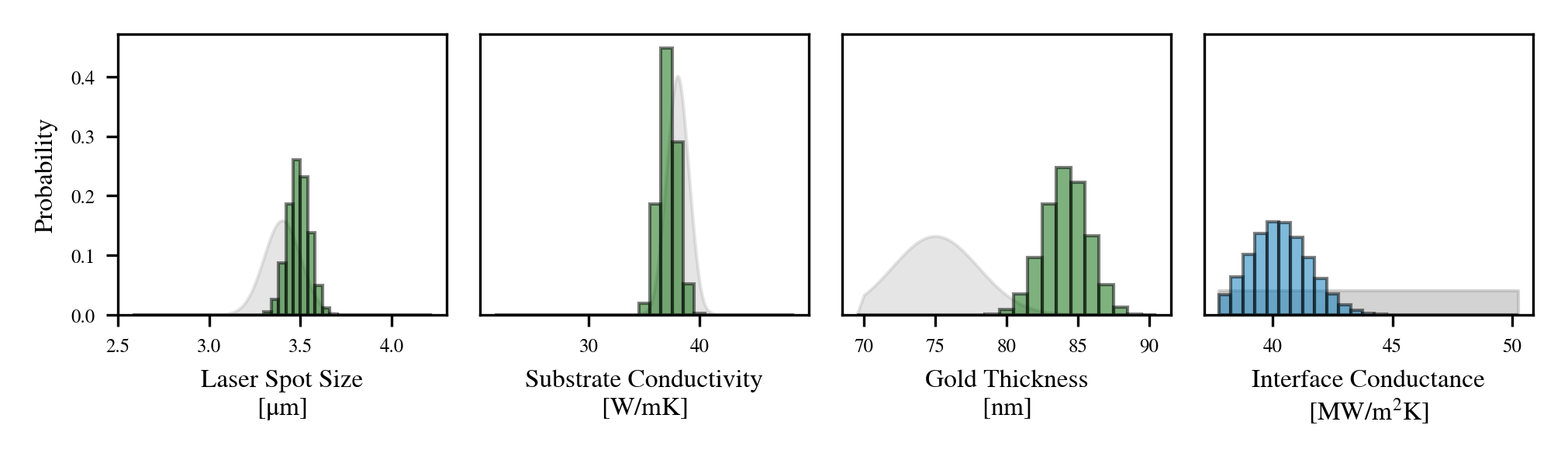}
    \caption{Model fits obtained using a Gaussian prior, with the original (incorrect) externally defined value of gold thickness (75 nm). Gray curves denote the associated prior
used for the externally defined input parameters, while green and blue bars signify the probabilities of the resulting inferences.}
    \label{fig:goldthickness}
\end{figure*}

Another notable example of this capability is the case of an incorrect assumption for an input parameter, also discussed in Sec.~\ref{sec:biased_prior} for the case of the thickness of the gold layer, which was initially assumed by line up with the deposition setpoint of 75 nm, while it was in reality substantially thicker (by more than 3$\times$ the associated uncertainty. Fig.~\ref{fig:goldthickness} shows the BPE results before this error was corrected (compare to Fig.~\ref{fig:different_priors} in the main text). It can be seen that even with a prior centered at 75 nm, BPE analysis of the FDTR measurements still signaled that the true gold thickness was higher. This prompted subsequent profilometry measurements, revealing the thickness to be $82\pm3$ nm.

\subsection{Additional Results of the Gaussian Prior}

\begin{figure}
        \centering
        \includegraphics[width=\linewidth]{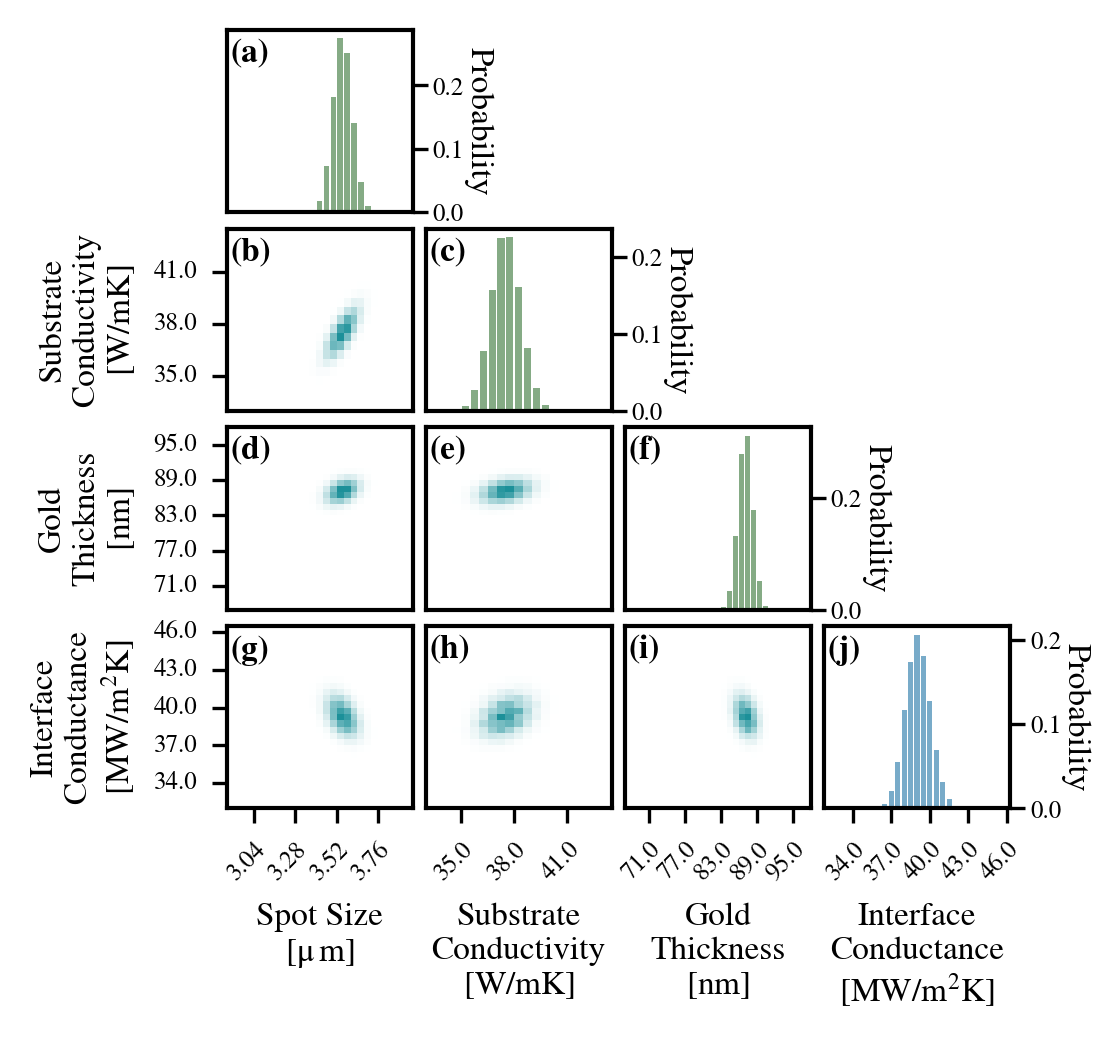}
        \caption{Resulting probabilities from the multi-parameter
inference with a Gaussian prior. This can be thought of as the result of Fig. \ref{fig:pairplot} when multiplied by the Gaussian weighting factors depicted in Fig. \ref{fig:different_priors}b. The four on-diagonal plots (a,
c, f, j) represent the marginalized probability distributions of
each individual parameter, with the y-axis representing the
relative probability that any individual parameter value is
correct. The six off-diagonal plots (b, d-e, g-i) represent
probability heatmaps of each pairwise combination of
parameters, showing the most likely pairwise combinations
of the parameter values.}
        \label{fig:corner_plot_gaussian_prior}
    \end{figure}

The Gaussian prior does not only affect the single-variable probability distributions that are depicted in Fig. \ref{fig:different_priors}. Rather, the weighting factors are implemented over the entire multi-parameter probability landscape, and so also affect the correlation structures that are observed between parameters, which had previously been visualized in Fig. \ref{fig:pairplot} for a uniform prior. The analogous plot for the Gaussian prior is shown in Fig. \ref{fig:corner_plot_gaussian_prior}.

Because the prior is effectively a weighting factor applied to the probability landscape, the correlations between variables should behave similarly, and indeed this appears to be the case when comparing the plots associated with the two priors. However, in cases where there are sets of solutions with similarly good fit, the Gaussian prior can constrain the high-probability regions to a smaller range of options. This behavior is observed in the off-diagonal plots in Fig \ref{fig:corner_plot_gaussian_prior}.

\begin{figure}
    \centering
    \includegraphics[width=1\linewidth]{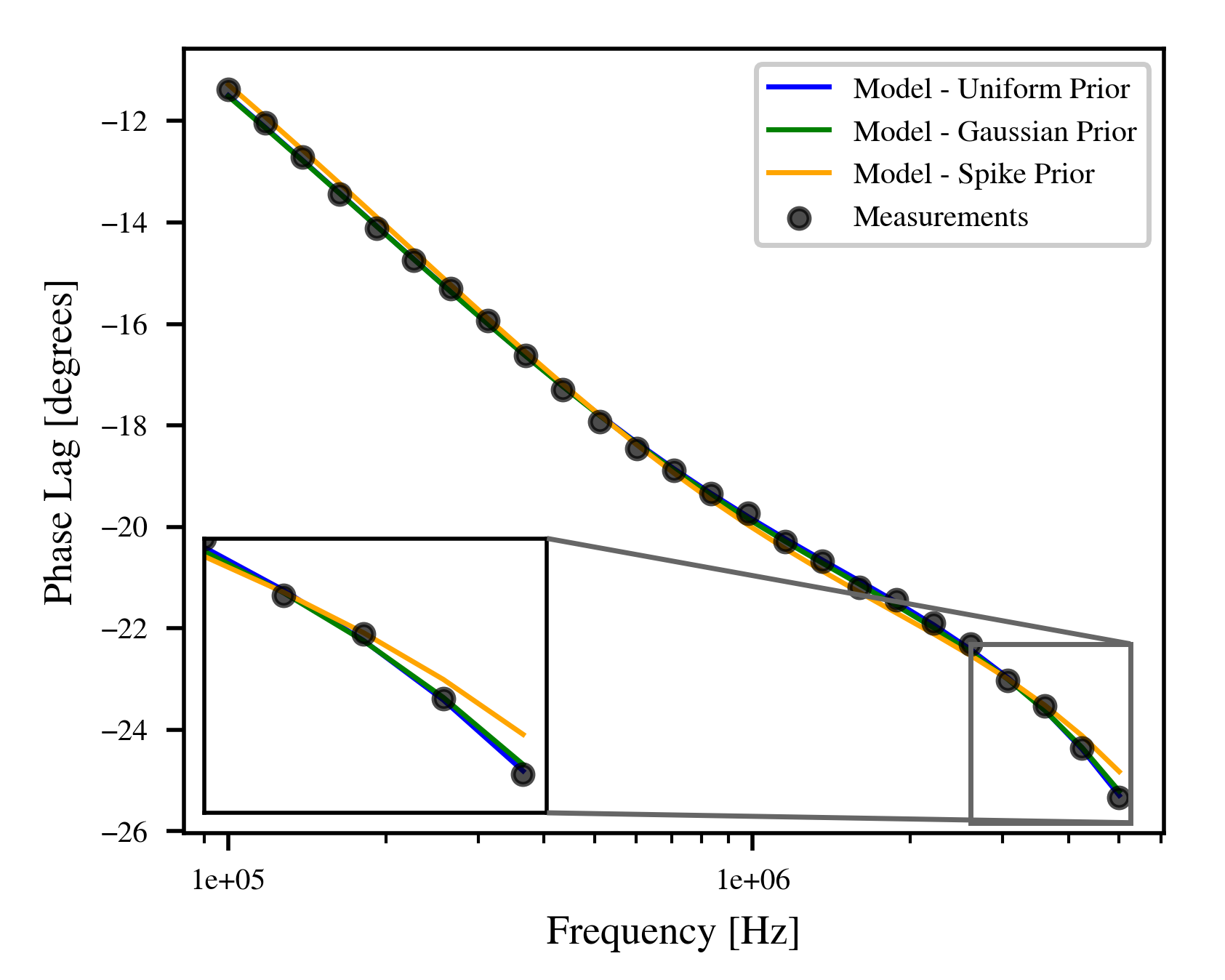}
    \caption{Phase lag measurements compared to the best-fit models obtained with uniform, Gaussian, and spike priors. The inset highlights the high-frequency region of the measurements, which contained one of the regions of highest discrepancy between the measurements and the results of the spike prior.}
    \label{fig:Unif_vs_Gauss_prior_phase_lag}
\end{figure}

In addition, it can be useful to examine the model fit that results from the different priors, as compared to the experimental measurements. This is shown in Fig. \ref{fig:Unif_vs_Gauss_prior_phase_lag}. Qualitatively, the model predictions associated with the different priors can be difficult to distinguish, although there are a few regions where their deviation is most significant. One of these regions is shown in the inset, where at high frequencies the spike prior is seen to deviate from the measurements, suggesting that one of our externally defined values may be slightly incorrect. In addition, the tight agreement between the uniform and Gaussian prior throughout the range of measurements can add confidence to our belief that constraining our externally defined inputs to the physically reasonable ranges with the Gaussian prior does not greatly reduce the quality of the fit.

\subsection{MSE Mapping}

\begin{figure}
    \centering
    \includegraphics[width=0.98\linewidth]{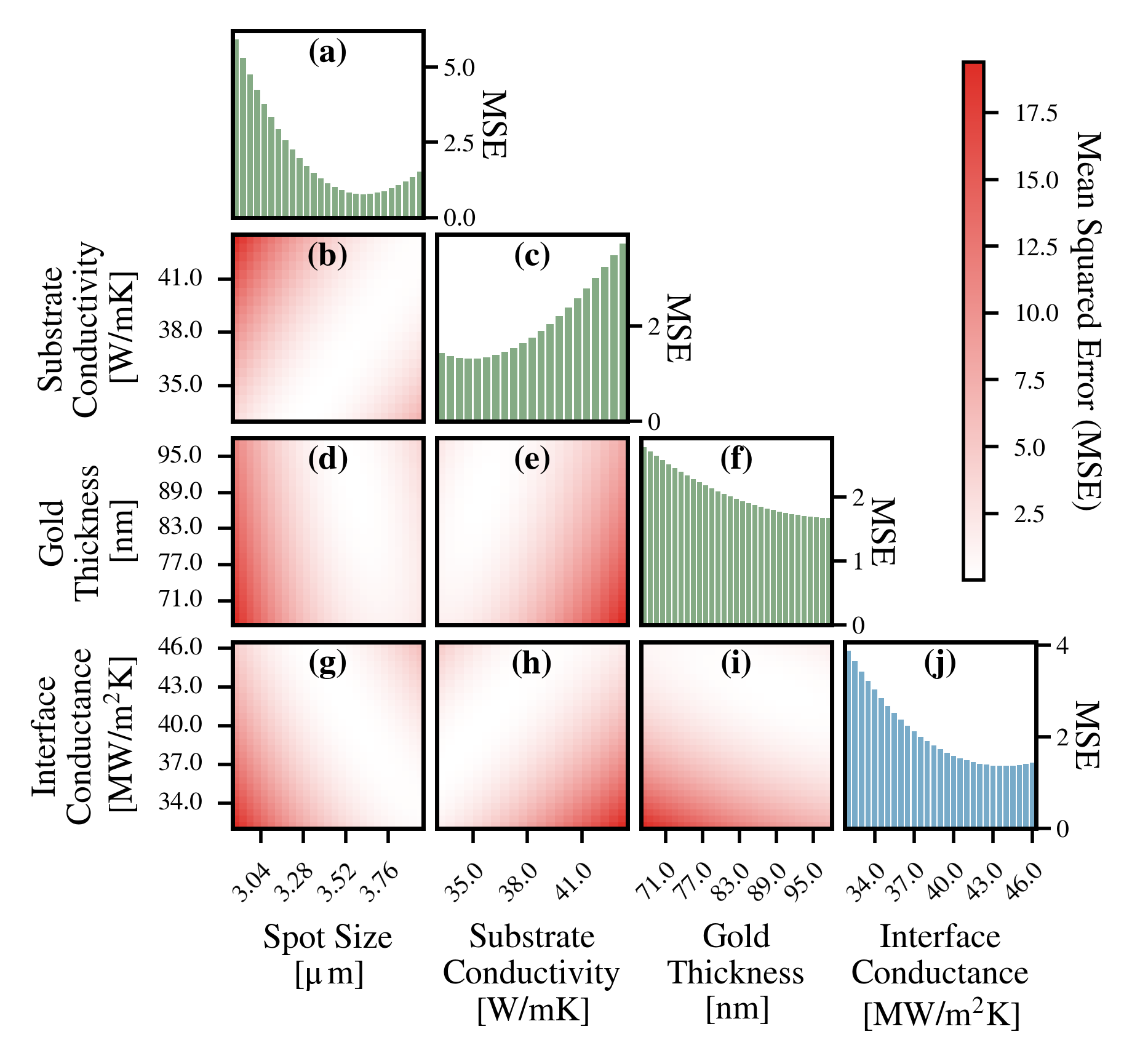}
    \caption{MSE mapping comparing the measured FDTR phase lags to the model predictions over the space of the 4 investigated parameters.  }
    \label{fig:4d_mse_map}
\end{figure}

MSE maps have been noted throughout this work as useful for visualizing the correlation between different parameters, and an MSE map of the relevant parameter ranges for ths work is shown in Fig. \ref{fig:4d_mse_map}. The interpretation can be thought of as the reverse of the probability landscapes in Figs. \ref{fig:pairplot} and \ref{fig:corner_plot_gaussian_prior}, in that the regions of lowest MSE (as opposed to highest probability) should correspond to closest agreement between the model and the experimental measurements.

It is somewhat natural to assume that, after accounting for this difference in representation, the probability landscape should very closely match the MSE map. In the case of uniform uncertainty for all experimental measurements, this should indeed be true for the set of probabilities that exists at each unique combination of values in the multi-parameter space. However, when visualizing high-dimensional data in 2D as in Figs. \ref{fig:pairplot}, \ref{fig:corner_plot_gaussian_prior}, and \ref{fig:4d_mse_map}, slight discrepancies can emerge in the visualizations. 

This results from the aggregation of values, and the fact that the probabilities scale down exponentially with the level of error, while the magnitude of the errors themselves are additive. Any asymmetry in error across one of the dimensions of the parameter space will result in an even more exaggerated asymmetry in probabilities, and the resulting marginalized plots in 1D and 2D will not match exactly. This discrepancy can in fact be an additional argument for the usefulness of probability landscapes, as it can be more valuable to identify a very close agreement with experiment at a specific set of parameter values, as opposed to a moderate level of agreement with experiment that holds over a wide range of parameter values.  

\subsection{Monte Carlo with MSE Cutoff} 

\begin{figure}
    \centering
    \includegraphics[width=\linewidth]{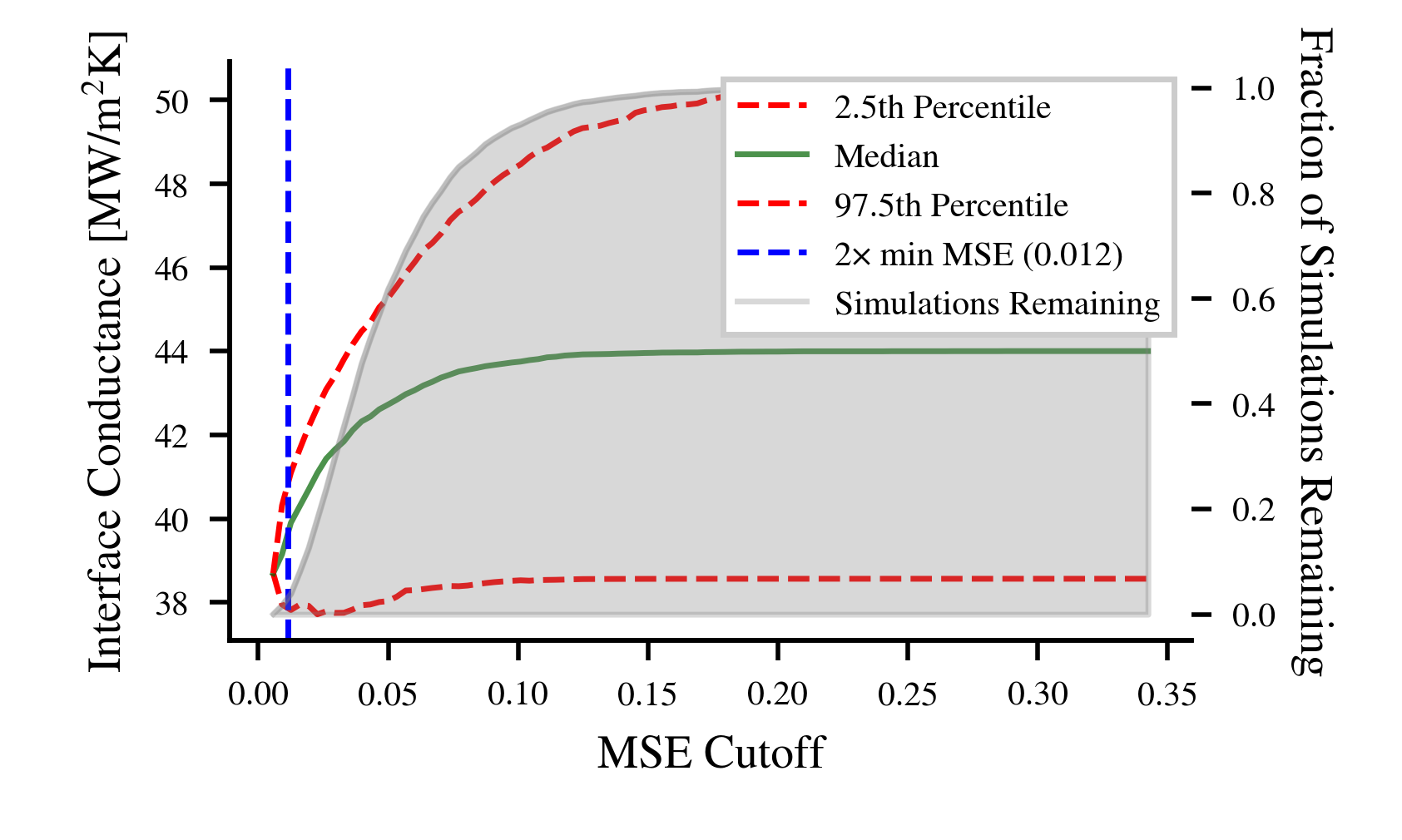}
    
        \caption{The resulting MC distribution of interface thermal conductance is plotted for a range of possible thresholds for MSE cutoff, with any simulation exceeding the cutoff being discarded. The 2.5th, 50th, and 97.5th percentile values of the interface thermal conductance are plotted, with the median shown by a solid green line and the 2.5th and 97.5th percentile values represented by dashed red lines. The gray curve represents the fraction of the original 3000 simulations that remain, as the MSE cutoff becomes more stringent. The dashed blue vertical line corresponds to the position equal to 2$\times$ the minimum MSE out of all 3000 simulations, the common rule-of-thumb threshold which has been discussed throughout this work. }
    \label{fig:mse-cutoff_MC}
\end{figure}

As discussed in Sec. \ref{mc_results}, an MSE threshold of varying magnitude was imposed on the original set of 3000 MC simulations, in order to confirm that fit-weighting was the major reason for the difference between BPE and MC results (since both methods explore the range of the parameter space). The results of this investigation were first shown in Fig. \ref{fig:all_bar_chart}, and are shown in more detail in Fig. \ref{fig:mse-cutoff_MC}.

It is implicit that lowering the MSE cutoff will reduce the number of simulations remaining in the aggregate, and this is shown by the gray curve in the background of the plot. Because MC methods rely on the random generation of simulation inputs according to the normal distributions shown in the left plots of Fig. \ref{fig:mc}, when there is a large number of simulations, most should have model input parameters at or near their externally defined values. However, as the MSE cutoff becomes more stringent, only the best-fitting simulations will remain, regardless of their proximity to the externally defined values. 

The fact that application of this MSE cutoff corresponds with a significant change in the median value and associated uncertainty range suggests that the externally defined values do not provide the best agreement with the experimental measurements. The MC distribution that remains at 2$\times$ the minimum MSE (the cross section corresponding to the dashed blue vertical line) has a similar median and uncertainty estimate to the BPE results, suggesting that the fit-weighting according to MSE is indeed the major source of discrepancy between MC and BPE methods.  

\end{document}

%% file: main.bbl
%